\documentclass[12pt,preprint]{aastex}
\usepackage{epsf}
\def\himsun{{h^{-1}M_\odot}}
\def\msun{{M_\odot}}
                                             
\def\pppm{\rm P^3M}
\def\mpc{\,h^{-1}{\rm {Mpc}}}
\def\kpc{\,h^{-1}{\rm {kpc}}}

\shorttitle{Triaxial Modeling of Halo Density Profiles}
\shortauthors{Jing \& Suto}
\begin{document}
%
%%%%%%%%%%%%%%%%%%%%%%%%%%%%%%%%%%%%%%%%%%%%%%%%%%%%%%%%%%%%%%%%%%%%%%
\title{Triaxial Modeling of Halo Density Profiles with
High-resolution N-body Simulations. }
%%%%%%%%%%%%%%%%%%%%%%%%%%%%%%%%%%%%%%%%%%%%%%%%%%%%%%%%%%%%%%%%%%%%%%
\author{Y.P. Jing} 
\affil{Shanghai Astronomical Observatory, the Partner Group of MPI f\"ur
Astrophysik, \\Nandan Road 80,  Shanghai 200030, China}
\email{ypjing@center.shao.ac.cn}
\and
\author{Yasushi Suto} 
\affil{Department of Physics and Research Center for the Early Universe
(RESCEU)\\ School of Science, University of Tokyo, Tokyo 113-0033,
Japan.}
\email{suto@phys.s.u-tokyo.ac.jp}
%%%%%%%%%%%%%%%%%%%%%%%%%%%%%%%%%%%%%%%%%%%%%%%%%%%%%%%%%%%%%%%%%%%%%%
%
\received{2002 February 5}
\accepted{2002 ???}
\begin{abstract}
We present a detailed non-spherical modeling of dark matter halos on
the basis of a combined analysis of the high-resolution halo
simulations (12 halos with $N\sim 10^6$ particles within their virial
radius) and the large cosmological simulations (5 realizations with
$N=512^3$ particles in a $100h^{-1}$Mpc boxsize).  The density
profiles of those simulated halos are well approximated by a sequence
of the concentric triaxial distribution with their axis directions
being fairly aligned.  We characterize the triaxial model
quantitatively by generalizing the universal density profile which has
previously been discussed only in the framework of the spherical
model. We obtain a series of practically useful fitting formulae in
applying the triaxial model; the mass and redshift dependence of the
axis ratio, the mean of the concentration parameter, and the
probability distribution functions of the the axis ratio and the
concentration parameter. These accurate fitting formulae form a
complete description of the triaxial density profiles of halos in Cold
Dark Matter models. Our current description of the dark halos will be
particularly useful in predicting a variety of nonsphericity effects,
to a reasonably reliable degree, including the weak and strong lens
statistics, the orbital evolution of galactic satellites and
triaxiality of galactic halos, and the non-linear clustering of dark
matter.  In addition, this provides a useful framework for the
non-spherical modeling of the intra-cluster gas, which is crucial in
discussing the gas and temperature profiles of X-ray clusters and the
Hubble constant estimated via the Sunyaev -- Zel'dovich effect.

\end{abstract} 
%
%%%%%%%%%%%%%%%%%%%%%%%%%%%%%%%%%%%%%%%%%%%%%%%%%%%%%%%%%%%
\section{Introduction}
%%%%%%%%%%%%%%%%%%%%%%%%%%%%%%%%%%%%%%%%%%%%%%%%%%%%%%%%%%%

The density profiles of dark matter halos have attracted a lot of
attention recently since \citet[NFW hereafter]{nfw96,nfw97} discovered
the unexpected scaling behavior in their simulated halos. Subsequent
independent higher-resolution simulations
\citep[e.g.,][]{fukushige97,fukushige01,moore98,jing00,jingsuto00}
confirmed the validity of the NFW modeling, in particular the presence
of the central cusp, although the inner slope of the cusp seems somewhat
steeper than they originally claimed.  Those previous models, however,
have been based on the spherical average of the density
profiles. Actually it is also surprising that the fairly accurate
scaling relation applies after the spherical average despite the fact
that the departure from the spherical symmetry is quite visible in
almost all simulated halos \citep[e.g., Fig.1 of ][]{jingsuto00}.

A more realistic modeling of dark matter halos beyond the spherical
approximation is important in understanding various observed properties
of galaxy clusters and non-linear clustering (especially the high-order
clustering statistics) of dark matter in general. In particular, the
non-sphericity of dark halos is supposed to play a central role in the
X-ray morphologies of clusters \citep{jing95, bx97}, in the cosmological
parameter determination via the Sunyaev-Zel'dovich effect
\citep{BHA91,inagaki95, yoshikawa98} and in the prediction of the
cluster weak lensing and the gravitational arc
statistics\citep{bart98,meneghetti00,meneghetti01,molikawa01,oguri01,keeton01}.
Nevertheless useful analytical modeling of the non-sphericity is almost
impossible, and numerical simulations are the only practical means to
provide statistical information.  

While the non-sphericity of the dark matter halos is a poorly studied
topic, some seminal studies do exist which attempt to detect and
characterize the non-spherical signature \citep[e.g.,][]
{BE87,Warren92,dubinski94,jing95,thomas98,yoshida00,meneghetti01,bullock01b}.
Nevertheless there is no systematic and statistical study to model and
characterize the density profiles of simulated halos. This is exactly
what we will present in the rest of the paper.  In particular, much
higher mass and spatial resolutions of our current N-body simulations
enable us to characterize the statistics of the halo non-sphericity
with an unprecedented precision.

This paper is organized as follows; two different sets of N-body
simulations that we extensively analyze here are described in \S 2.  In
\S 3, we discuss how to define the iso-density surfaces of dark mater
halos from simulation data and then argue that they are well
approximated by a sequence of the concentric triaxial model. Section 4
characterizes the statistical distribution of the triaxial model
parameters. Finally \S 5 is devoted to summary and discussion.

%%%%%%%%%%%%%%%%%%%%%%%%%%%%%%%%%%%%%%%%%%%%%%%%%%%%%%%%%%%
\section{Simulations for  dark matter halos}
%%%%%%%%%%%%%%%%%%%%%%%%%%%%%%%%%%%%%%%%%%%%%%%%%%%%%%%%%%%

We use two different simulations for the current purpose. The first is
our new set of cosmological N-body simulations with $N=512^3$
particles in a $100h^{-1}$Mpc box, and the other is a set of
high-resolution halo simulation runs. We describe the two simulations
in the next subsections in order.

\subsection{Cosmological simulations}

The first set of simulations is our new runs with $N=512^3$ particles in
a $100h^{-1}$Mpc box.  These runs have been carried out in 2001 with our
Particle-Particle-Particle-Mesh ($\pppm$) code on the vector-parallel
machine VPP5000 at the National Astronomical Observatory of Japan. The
code adopts the standard $\pppm$ algorithm \citep{HE1981,efstathiou85},
is vectorized \citep{JS1998}, and has been recently parallelized. A mesh
of $1200^3$ grid points is used for the Particle-Mesh (PM) force
computation with the optimized Green function \citep{HE1981}. The
short-range force is compensated for the PM force calculation at the
separation less than $\epsilon=2.7H$, where $H$ is the mesh cell size
\citep{efstathiou85}. The linked-list technique has been used for
computing the short-range Particle-Particle (PP) interaction with
$448^3$ linked-list cells.  The computer has a total of 64 processors,
and we use $N_{\rm CPU} =8$ to 32 processors, upon their availability,
to run our code. The most important advantage of the machine for our
work is that each processor has a big memory of $16{\rm GB}$, sufficient
for storing all the information of the code. The PM computation can be
easily parallelized, and it is crucial to parallelize the PP computation
which dominates the CPU computation time for a strongly clustered
simulation like our present case. We sliced the simulation box in one
direction (e.g. z-axis) with the thickness chosen to be the cell size of
the linked-list cell. Those 448 slices in total are sorted in the
descending order according to the number of particles they contain. We
distribute the PP force computation among the different processors in a
simple way; $n$-th processor ($n=1$ to $N_{\rm CPU}$) is assigned the
force computation for those slices with indices of $j N_{\rm CPU}+n$
where $j$ runs from 0 to $j_{\rm max} \equiv 448/N_{\rm CPU} -1$.  The
PP interaction of the particles in the same slice and in the adjacent
lower slice is considered, so the interaction for each pair of particles
is computed only once.  With this computation partition, we find that
the load-balance problem, which becomes progressively serious for
$\pppm$ simulations in the later strongly clustered regime, can be
overcome to a satisfactory degree; even at the final stage of our
simulation runs, the CPU time for the PP part is nearly inversely
proportional to the number of the processors used. This implies that the
code has achieved a good parallelization efficiency.

We consider two representative cold dark matter (CDM) models; a
low-density flat cosmological model (LCDM) with $\Omega_0=0.3$ and
$\lambda_0=0.7$, and the Einstein-de Sitter model with $\Omega_0=1$
(SCDM). The primordial density fluctuation is assumed to obey the
Gaussian statistics, and the power spectrum is given by the
Harrison-Zel'dovich type. The linear transfer function for the dark
matter power spectrum is taken from \citet{BBKS1986}.  The shape and
the normalization of the linear power spectrum are specified by the
shape parameter, $\Gamma=\Omega_0 h$, and $\sigma_8$ respectively,
where $h$ is the Hubble constant in $100 {\rm km s^{-1} Mpc^{-1}}$ and
$\sigma_8$ is the {\it rms} linear density fluctuation within the
sphere of the radius $8\mpc$.  Table~\ref{table:cosmosim} summarizes
the physical and simulation parameters used for these simulations.  We
adopted $\sigma_8=0.9$ for LCDM and 0.55 for SCDM, both of which are
slightly smaller than those in our previous simulations
\citep{JS1998}, but seem more consistent with recent observations
(e.g., Seljak 2002; Lahav et al. 2002).  With the adopted values for
those physical parameters, the LCDM model satisfies almost all current
observations while the SCDM model is known to have many difficulties.
Therefore we mainly analyze the LCDM model for our purpose, and
sometimes use the SCDM simulation just for comparison.
  
The boxsize of our cosmological simulations is $100\mpc$, so the
particle mass is $m_p= 6.2\times 10^{8}\himsun$ and $2.1\times
10^{9}\himsun$, respectively, for the LCDM and SCDM simulations (Table
\ref{table:cosmosim}). The force resolution is $\eta=20\kpc$ for the
linear density softening form (Efstathiou et al. 1985; this roughly
corresponds to $\eta/3$ for the Plummer-type softening length).  The
simulations are evolved by 1200 time steps from the initial redshift
$z_i=72$.  Two realizations are computed for each model. One additional
LCDM simulation (LCDMa) uses a smaller force softening $\eta=10\kpc$ and
is evolved with 5000 time steps in order to check the possible effect of
the force softening on the final dark matter distribution especially at
small scales.  As far as the shape of the virialized halos is concerned,
we made sure that both simulations (LCDM and LCDMa) yield almost
identical results.  In what follows, therefore, we do not distinguish
LCDM and LCDMa, and simply refer to them as LCDM.

\subsection{Identification of dark halos in the cosmological simulations}

The Friends-Of-Friends (FOF) method is a widely used algorithm to
identify dark matter clumps in N-body data. The mean overdensity
within the clumps is approximately proportional to $b^{-3}$, where $b$
is the bonding length. It has been shown that the FOF clumps with
$b=0.2 {\bar d}$, where $\bar d \equiv L/N^{1/3}$ is the mean
separation of particles, approximately correspond to the virialized
dark matter halos of the mean overdensity 180
\citep[e.g.,][]{davis85,LC94}.  On the other hand, a large fraction of
the FOF clumps identified with $b=0.2 {\bar d}$ are known to form a
system of multiple virialized halos that are bridged via thin
filaments (e.g., Suto, Cen \& Ostriker 1992; Suginohara \& Suto 1992;
Jing \& Fang 1994, hereafter JF94). JF94 proposed to compute the
overdensity around the local potential minima within each FOF clump to
separate the virialized halos. While this can effectively achieve the
goal, it is time-consuming to find the local potential minima (because
there may be multiple minima within a single FOF clump).

Here we propose to use an alternative method which works faster. The
thin bridges connecting the halos identified with $b=0.2 {\bar d}$ can
be effectively eliminated by reducing $b$. By trial and test, we found
that the thin bridges almost disappear if we adopt $b=0.1 {\bar
d}$. With this recipe, however, the resulting FOF clumps have a smaller
size and a higher overdensity than those defined according to the
spherical collapse model.  Therefore our scheme should be interpreted to
identify first the central parts or the substructures of the entire
halo. Next, for each FOF clump of $b=0.1 {\bar d}$, we compute the
gravitational potential of every member particle. The position of the
particle of the minimum potential is defined as the center of the
hosting halo. Then the spherical overdensity is computed around the halo
center with increasing the radius, and the virial radius $r_{\rm vir}$
is found when the overdensity reaches the value predicted in the
spherical collapse model. Here we use the fitting formula of
\citet{BN98} for the spatially-flat ($\Omega(z)+\lambda(z)=1$) models:
%%%%%%%%%%%%%%%%%%%%%%%%%%%%%%%%%%%%%%%%%%%%%%%%%%%%%%%%%%%%%%%%%%%
\begin{equation} 
\label{eq:virial}
\Delta_{\rm vir}(z) \equiv \frac{3 M_{\rm vir}}{4\pi r^{3}_{\rm vir}
\rho_{crit}} =18\pi^2 + 82
\left[\Omega(z)-1\right]-39\left[\Omega(z)-1\right]^2 ,
\end{equation} 
%%%%%%%%%%%%%%%%%%%%%%%%%%%%%%%%%%%%%%%%%%%%%%%%%%%%%%%%%%%%%%%%%%%
where $\rho_{crit}$ is the critical density of the Universe.  Since
our choice $b=0.1 {\bar d}$ preferentially selects smaller clumps than
those predicted in the spherical model, some fraction of such
clumps turn out to be substructures within the virial radius of a
larger halo defined in the above equation. If the virial spheres of
more than one halos overlap, we simply retain the most massive clump
and throw away the others from the final halo list.

\subsection{High-resolution halo simulations}

Our cosmological simulations which we described above have a
sufficient spatial resolution to discuss the statistics concerning the
halo shapes and the concentration of the density profile (\S 4 and 5)
as was conducted by Jing (2000) in the framework of the spherical
approximation.  Actually except for a delicate problem of determining
the slope of the central cusp at $r\ll 0.01 r_{vir}$, a larger
simulation volume is more important than the higher resolution for the
current purpose.  Nevertheless we also use our higher-resolution halo
simulations (Jing \& Suto 2000; hereafter simply referred to halo
simulations) to demonstrate that our triaxial modeling indeed provides
a better description for halo profiles than the conventional spherical
modeling (\S 3).

These halos are simulated with about a million particles within their
virial radii (see Table 1 of Jing \& Suto 2000).  For mass scales of
clusters, groups, and galaxies, there are four halos, respectively, and
thus twelve halos in total.  They are simulated in the LCDM model except
the fact that the fluctuation amplitude, $\sigma_8=1$ (Kitayama \& Suto
1997), is a bit larger than our current choice $\sigma_8=0.9$.  Another
advantage of the halo simulations is that those halos are simulated with
almost the equal number of particles independently of the mass of the
halos, and thus the resolution relative to the virial radius and the
halo mass is kept constant. This is not the case for the cosmological
simulations in which massive halos would have a better resolution in
terms of the number of particles involved. Thus the possible artificial
effect due to the variable resolution is suppressed in the halo
simulations.

After Jing \& Suto (2000) was published, we have completed runs of
additional two halos with a galactic mass and with a group mass. Those
new halos are referred to GX5 and GR5, respectively, according to our
previous convention. While we add these two, we also eliminate two
previous halos from the list of halos that we examine below; GR2 which
shows a clear bi-modal structure, and GX1 which is seriously disrupted
at $z\le 0.5$ due to the tidal force of a nearby massive object. This
is because the major purpose of analyzing the halo simulation
catalogs is to check the validity of the triaxial modeling for
typical halos. The fraction of those atypical halos is properly taken
into account in the statistics drawn from the cosmological
simulations. Thus the above replacement does not bias our conclusion.

%%%%%%%%%%%%%%%%%%%%%%%%%%%%%%%%%%%%%%%%%%%%%%%%%%%%%%%%%%%%%%%%%%%%%%%%%%%
\section{Modeling the non-spherical density profiles of dark matter halos}
%%%%%%%%%%%%%%%%%%%%%%%%%%%%%%%%%%%%%%%%%%%%%%%%%%%%%%%%%%%%%%%%%%%%%%%%%%%

In this section, we propose that a non-sphericity in the density
profiles of dark halos is well described by a triaxial model on the
basis of the detailed analysis of the halo simulations.  In fact, we
demonstrate that the triaxial modeling significantly improves the fit
to the simulated profiles, at least for relatively relaxed halos,
compared to the conventional spherical model.  The statistical
description including the probability distribution functions for axis
ratios and the concentration parameters will be discussed in the next
section using the cosmological simulations.

\subsection{Defining the iso-density surfaces inside individual halos}

The shapes of dark halos have been previously studied by many authors
\citep[e.g., ][]{BE87,Warren92,jing95,thomas98}, and it is already
well known that they exhibit a significant amount of departure from
spherical distribution.  Those previous studies first compute the
inertial tensor for each halo, and then compute the distribution of
the axial ratios and the correlation of the direction of the principal
axes.  While this is a well-defined method to characterize the shape
of halos in principle, we do not employ this for two reasons.

First, this method assumes that we know {\it in advance} which particles
belong to each halo.  In reality this is not the case since we usually
attempt to determine the member particles of a halo and its shape
simultaneously.  This is serious because the inertia tensor is sensitive
to the outer boundary of the halo where the membership of particles is
also difficult to define.  Previous studies get around the problem by
applying the procedure iteratively; first, all particles within a
certain spherical radius from the center of halo are included to compute
the inertial tensor and the resulting ellipsoidal configuration.  Next,
those particles outside the ellipsoid are thrown away from the member
particles of the halo, and the inertia tensor is re-calculated.  This
procedure is repeated until the solution converges.  While this method
seems to work well in previous low-resolution N-body data, we were not
able to obtain a good convergence in the case of our high-resolution
halos. This is ascribed to the fact that our high-resolution halos
retain a significant amount of substructures which have been
artificially erased due to the {\it overmerging} effect in previous
lower-resolution simulations. The iteration procedure is not stable in
the presence of significant substructures especially at the boundary
region of halos, since the inertial tensor is quite sensitive to them.

Second, our main interest here is not simply to define the overall shape
of halos, but to characterize the density profile. Therefore we would
like to have a sequence of {\it iso-density} surfaces with different
overdensities. The ellipsoidal surface obtained from the above
procedure, even if it converges, is not related to those iso-density
surfaces, and thus not so useful after all for our purpose here.

With the above problems of the previous method in mind, we propose
another approach to find the iso-density surfaces.  This begins with the
computation of a local density at each particle's position.
We adopt the smoothing kernel widely employed in 
the Smoothed Particle Hydrodynamics (SPH) method 
\citep[e.g., ][]{HK89}:
%%%%%%%%%%%%%%%%%%%%%%%%%%%%%%%%%%%%%%%%%%%%%%%%%%%%%%%%%%%%%%%%%%%
\begin{equation}
\label{eq:kernel}
W(r,h_i)=\frac{1}{\pi h_i^3} = \left\{ 
\begin{array}{ll}
\displaystyle 1-\frac{3}{2}\left(\frac{r}{h_i}\right)^2 
+ \frac{3}{4}\left(\frac{r}{h_i}\right)^3 
& (r\le h_i) \\
\displaystyle \frac{1}{4} \left(2-\frac{r}{h_i}\right)^3 
& (h_i<r< 2h_i) , \\
0  & {\rm otherwise}
\end{array}
 \right. 
\end{equation}
%%%%%%%%%%%%%%%%%%%%%%%%%%%%%%%%%%%%%%%%%%%%%%%%%%%%%%%%%%%%%%%%%%%
where $h_i$ is the smoothing length for the $i$-th particle. We use 32
nearest neighbor particles to compute the local density $\rho_i$, and
$h_i$ is set to be a half of the radius of the sphere that contains
those 32 neighbors. Using $\rho_i$, we construct the iso-density
surfaces corresponding to the 5 different thresholds:
%%%%%%%%%%%%%%%%%%%%%%%%%%%%%%%%%%%%%%%%%%%%%%%%%%%%%%%%%%%%%%%%%%%
\begin{eqnarray}
\label{eq:rhos}
 \rho_{\rm s}^{(n)} &=& A^{(n)} \rho_{\rm crit} , \\
\label{eq:an}
A^{(n)} &=& 100 \times 5^{n-1} \quad (n=1 \sim 5) .
\end{eqnarray}
%%%%%%%%%%%%%%%%%%%%%%%%%%%%%%%%%%%%%%%%%%%%%%%%%%%%%%%%%%%%%%%%%%% 
In practice, we collect all particles satisfying $0.97 \rho_{\rm
s}^{(n)} <\rho_i < 1.03\rho_{\rm s}^{(n)}$ to define the $n$-th
isodensity surface.  The typical sizes (the mean radii) of those
surfaces are 0.6, 0.4, 0.25 0.12 and 0.06 times the virial radius of the
halo, respectively.  Note that $\rho_{\rm s}^{(n)}$ is the {\it
local} density, and thus the {\it mean} density of the halo inside the
corresponding radius of $\rho_{\rm s}^{(n)}$ is generally much higher.

Actually a straightforward application of equation (\ref{eq:rhos})
results in many small distinct regions with the identical density
threshold inside an individual halo.  This is again due to the
presence of the strong substructures in the halo.  Since we are
interested in the isodensity surfaces which represent the overall
density profile of the parent halo, we have to eliminate those small
regions corresponding to the substructures. For this purpose, we again
use the FOF technique but with a different bonding length from that we
used when identifying the virialized halos.  After some trial and
error, we find that an adaptive (i.e., dependent on each isodensity
value) bonding length of $b_n=3 (\rho_{\rm s}^{(n)}/m_p)^{-1/3}$ works
well \citep[c.f., ][]{SCO92}.

\subsection{Triaxial model fits to the iso-density surfaces}

\begin{figure}[htb]
\begin{center}
 \leavevmode\epsfxsize=6.0cm \epsfbox{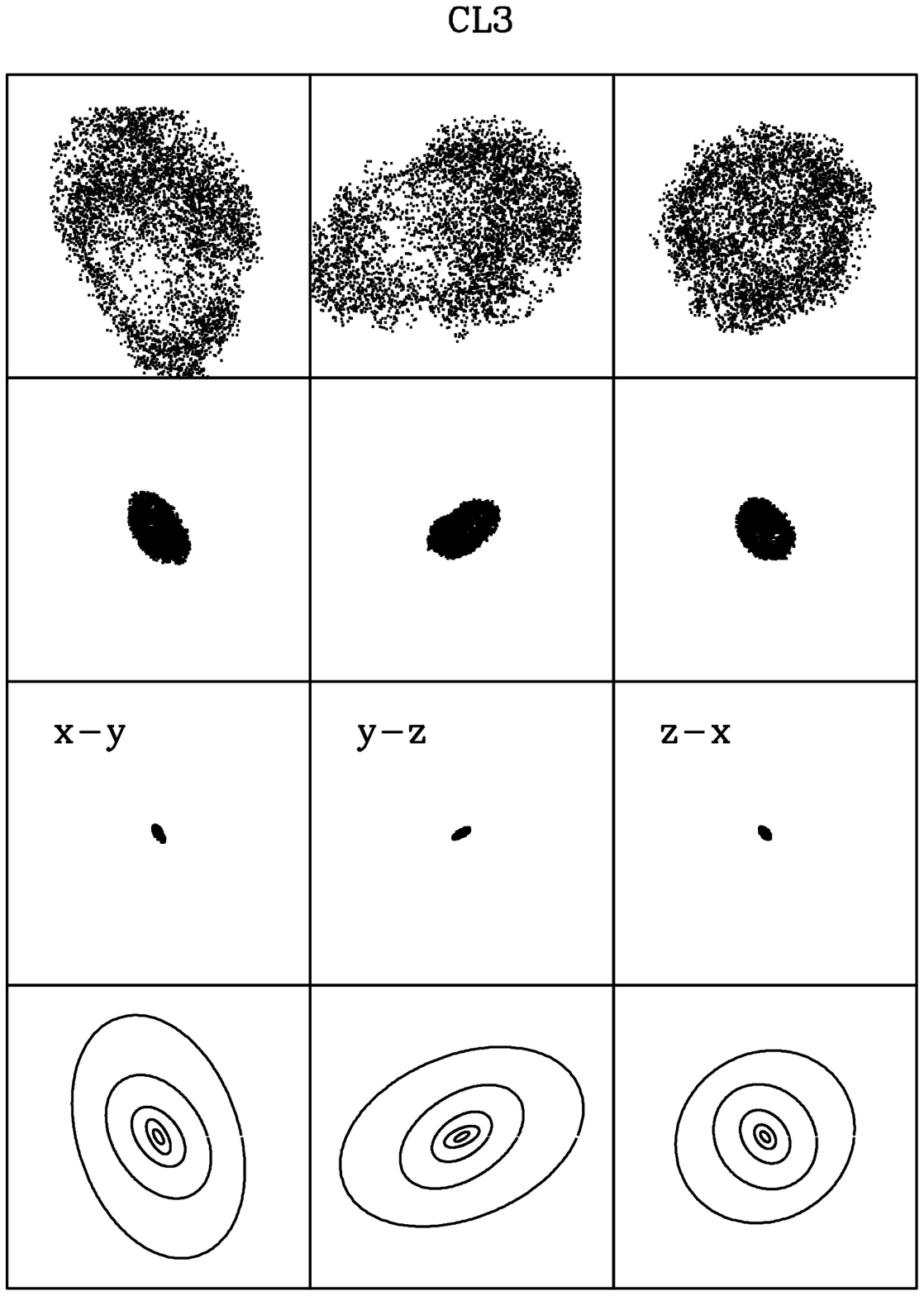}
 \leavevmode\epsfxsize=6.0cm \epsfbox{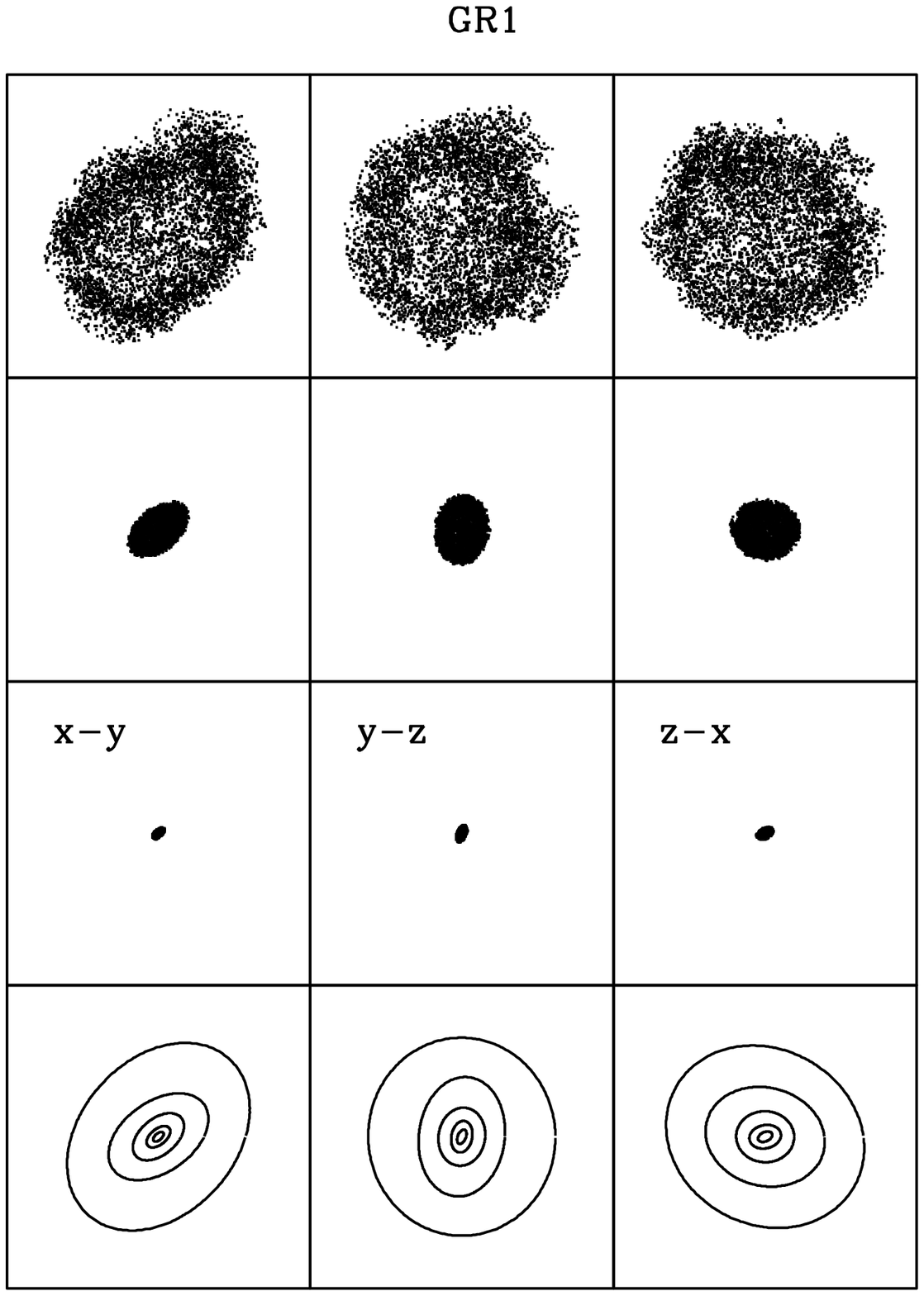}
\vspace*{0.5cm}
 \leavevmode\epsfxsize=6.0cm \epsfbox{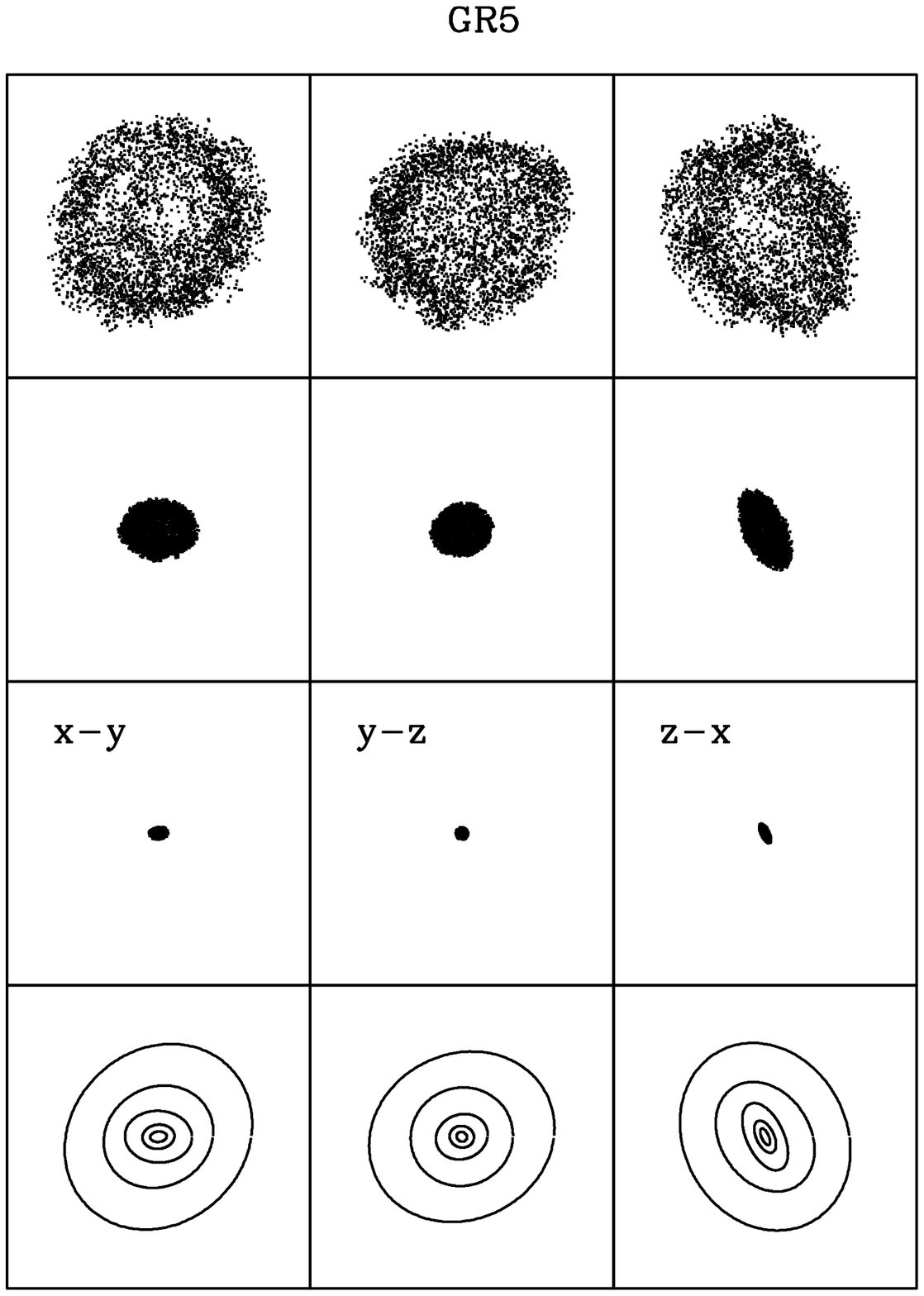}
 \leavevmode\epsfxsize=6.0cm \epsfbox{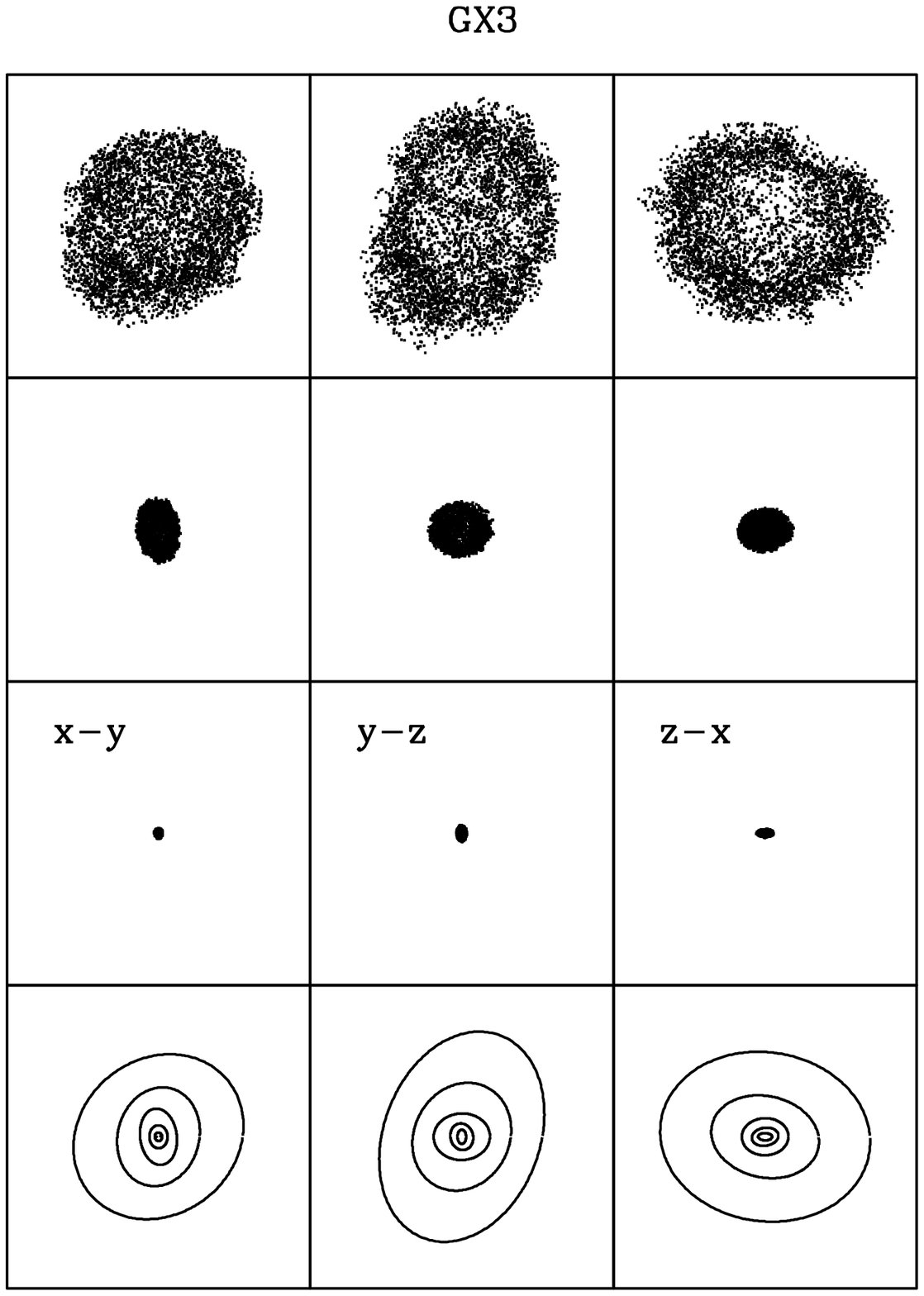}
\end{center} 
\figcaption{Examples of projected particle distribution in four halos;
 a) CL3, b) GR1, c) GR5, and d) GX3. The size of each box is $2r_{\rm
 vir}$ of each halo.  For each halo, particles in the isodensity shells
 with $A \equiv \rho_s/\rho_{\rm crit}=100$, $2500$, and $6.25\times
 10^4$ are plotted on the $xy$, $yz$ and $zx$ planes (from left to
 right). The bottom panels show the triaxial fits to five isodensity
 surfaces projected on those planes.  \label{fig:project} }
\end{figure}

\begin{figure}[htb]
\begin{center}
 \leavevmode\epsfxsize=8.0cm \epsfbox{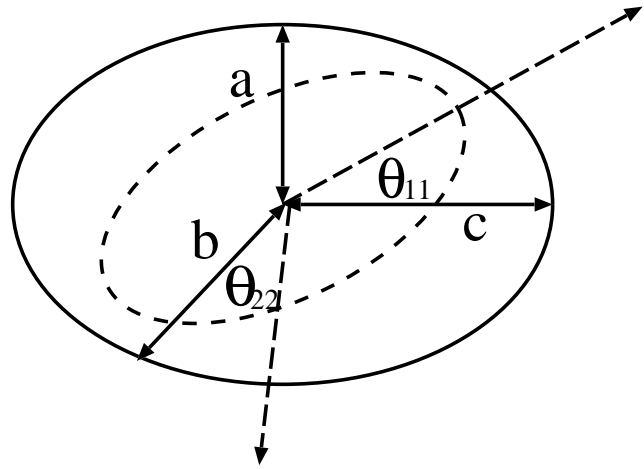}
\end{center} 
\figcaption{A schematic illustration of the triaxial model for the
 isodensity surface of dark matter halos. The axis lengths are defined
 to be $a \leq b \leq c$, and $\theta_{11}$ ($\theta_{22}$) measures the
 angle between the longest (middle) axis of the iso-density surface
 with that of $A^{(3)}=2500$ (eq.[\ref{eq:an}]).  \label{fig:haloshape} }
\end{figure}

Figure~\ref{fig:project} plots typical examples of the projected
particle distributions within the isodensity surfaces for four different
halos (CL3, GR1, GR5 and GX3) after particles in strong substructures
are eliminated as described above.  Those plots clearly suggest that the
isodensity surfaces are typically approximated as triaxial
ellipsoids. So we performed the following triaxial fit to the
iso-density surfaces with 5 different thresholds separately:
%%%%%%%%%%%%%%%%%%%%%%%%%%%%%%%%%%%%%%%%%%%%%%%%%%%%%%%%%%%%%%%%%%%
\begin{eqnarray}
\label{eq:triaxial}
 R^2(\rho_{\rm s}) = \frac{X^2}{a^2(\rho_{\rm s})} 
+ \frac{Y^2}{b^2(\rho_{\rm s})} + \frac{Z^2}{c^2(\rho_{\rm s})} .
\end{eqnarray}
%%%%%%%%%%%%%%%%%%%%%%%%%%%%%%%%%%%%%%%%%%%%%%%%%%%%%%%%%%%%%%%%%%%
The origin of the coordinates is always set at the center of mass of
each surface, and the principal vectors ${\bf a}$, ${\bf b}$ and ${\bf
c}$ ($a \le b \le c$) are computed by diagonalizing the inertial tensor
of particles in the surface (Fig.\ref{fig:haloshape}).  The projected
views of the corresponding fitted ellipsoids are shown at the bottom
panels in Figure~\ref{fig:project}, which implies that the ellipsoid
fitting is a good approximation (at least visually).

\begin{figure}[htb]
\begin{center}
 \leavevmode\epsfxsize=10.0cm \epsfbox{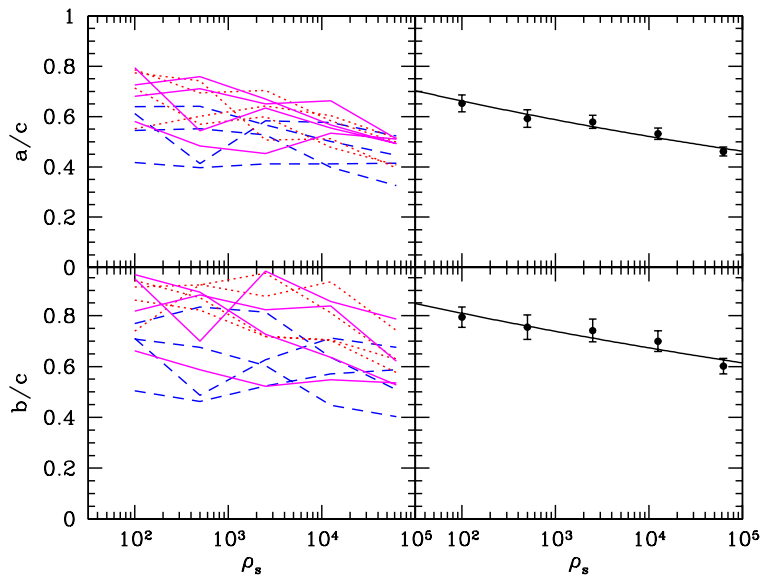}
\end{center} 
\figcaption{Axis ratios for the triaxial model fits to twelve halos.
{\it Left:} results for individual halos. The dashed lines are for
cluster halos, the dotted ones for group halos, and the solid lines for
galactic halos. {\it Right:} symbols indicate the mean and its one-sigma
error from the halo simulations, while the solid lines show the
single power-law fit (eq.[\ref{eq:axisratio}]). The upper and lower
panels show $a/c$ and $b/c$, respectively. \label{fig:axisratio} }
\end{figure}

Figure \ref{fig:axisratio} plots the dependence of the axis ratios,
$a/c$ and $b/c$, on the isodensity threshold $\rho_{\rm s}$.
Naturally each halo exhibits different behavior which may reflect the
different merging history and/or tidal force field. Nevertheless,
several systematic dependences are quite visible. The halos of cluster
mass generally have smaller axial ratios than those of galactic mass,
implying that the halos of the galactic mass are rounder on average
than those of cluster mass.  This mass-dependence will be quantified
with a large sample of halos from the cosmological simulations in \S
4.

On the other hand, we also note that the axial ratios decrease with
increasing the density; the iso-density surfaces become more elongated
in the central region than in the outer region. The mean (with the error
bar of the mean) of the
axial ratios computed from the twelve halos are plotted in the right
panels of Figure \ref{fig:axisratio} (the symbols).  The solid lines
show the single power-law fit for the mean axis ratios:
%%%%%%%%%%%%%%%%%%%%%%%%%%%%%%%%%%%%%%%%%%%%%%%%%%%%%%%%%%%%%%%%%%%
\begin{eqnarray}
\label{eq:axisratio}
\frac{a}{c} &=& 0.56
\left(\frac{\rho_{\rm s}/\rho_{\rm crit}}{2500}\right)^{-0.052} \\
\frac{b}{c} &=& 0.71
\left(\frac{\rho_{\rm s}/\rho_{\rm crit}}{2500}\right)^{-0.040} .
\end{eqnarray}
%%%%%%%%%%%%%%%%%%%%%%%%%%%%%%%%%%%%%%%%%%%%%%%%%%%%%%%%%%%%%%%%%%%

\begin{figure}[htb]
\begin{center}
 \leavevmode\epsfxsize=10.0cm \epsfbox{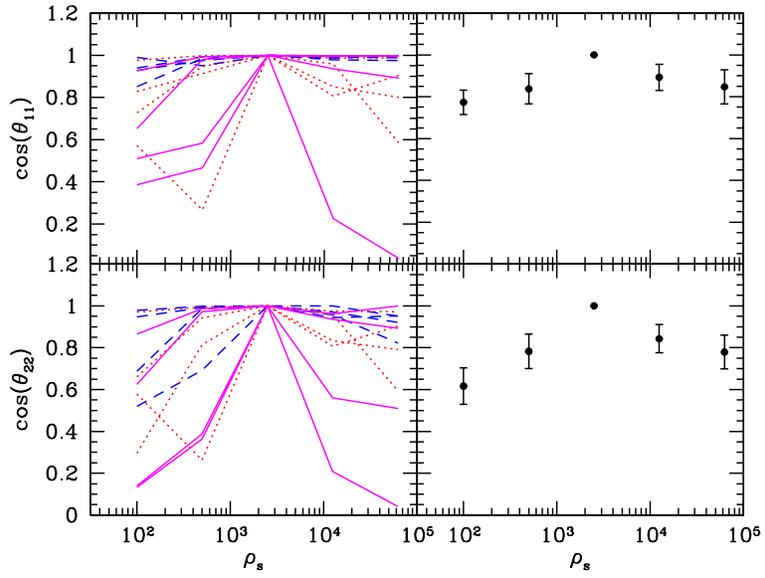}
\end{center} 
\figcaption{Degree of alignment of the directions of the ellipsoid axes.
{\it Left:} results for individual halos. The dashed lines are for
cluster halos, the dotted ones for group halos, and the solid lines for
galactic halos. {\it Right:} symbols indicate the mean and its one-sigma
error from the halo simulations.  The upper and lower panels show
for the major and middle axes respectively.  \label{fig:costheta} }
\end{figure}

Figure \ref{fig:costheta} shows the degree of the alignment of the axis
directions among isodensity surfaces at different densities (radii).  We
define $\theta_{11}$ as the angle between the major axis of the
isodensity surfaces and that of the $A^{(3)}=2500$ isodensity surface as
shown in Figure \ref{fig:haloshape}. Similarly, $\theta_{22}$ is defined
with respect to their middle axes.  According to our definition, $\cos
\theta_{11}= \cos \theta_{22}=1$ at $\rho_s/\rho_{\rm
crit}=A^{(3)}=2500$.

We find that the major axes align pretty well within a halo; for about
$70\%$ of the halos $\cos \theta_{11}$ at different radii is larger than
0.7. For about half of the sample, $\cos \theta_{11}$ is larger than
0.9. In a few cases (3 out of 12 halos), however, the alignment of the
major axes is poor. When we check these halos individually (e.g. GR1),
it turns out that $b/c$ for the two halos is quite close to unity,
indicating they are oblate halos with $b\approx c$ and thus the
direction of the major axis is difficult to measure (if $b=c$, the
direction of the major axis is arbitrary within a plane). Thus the
apparent mis-alignment of their major axes is not meaningful.  Only
for the remaining one halo (GX3; Figure~\ref{fig:project}), the major
axes of the outer and the innermost isodensity surfaces are indeed
perpendicular to that at the middle. This is the real case that the
major axes are significantly mis-aligned.

The alignments of the middle axes show similar behavior: for most of
the halos the degree of the alignment is satisfactory. For those which
show significant misalignment of the middle axes, their $a/b$ or $b/c$
ratio is usually quite close to unity and the direction of the middle
axes (and the minor or major axes) can be poorly determined at best.
Only in the case like GX3, no simple ellipsoid description can be
found, but this is fairly exceptional.  The alignment seems slightly
better for cluster-sized halos, but this would be simply because
galactic halos are more spherical and thus the direction of the major
axis is less accurate than that for cluster-sized halos.

\subsection{Triaxial versus spherical modeling of dark halos}

In the last subsection, we have seen that the isodensity ellipsoids at
different radii are approximately aligned, and the axial ratios of the
ellipsoids are nearly constant. These facts suggest the
possibility that the internal density distribution within a halo can
be approximated by a sequence of the concentric ellipsoids of a
constant axis ratio. To show this to be an improved description over
the conventional spherical description, we compute the quadrupole of
the particle distribution within a spherical shell ($Q_s$) or an
ellipsoid shell ($Q_e$). For a spherical shell, the positions of
particles inside the shell can be described by
%%%%%%%%%%%%%%%%%%%%%%%%%%%%%%%%%%
\begin{equation}
\left\{ 
\begin{array}{l}
x = r \sin \theta \cos \phi \cr
y = r \sin \theta \sin \phi \cr
z = r \cos \theta
\end{array}
 \right. 
\label{eq:sphcoord}
\end{equation}
%%%%%%%%%%%%%%%%%%%%%%%%%%%%%%%%%%
with $r$ being the (conventional) spherical radius.  Similarly, the
positions of the particles in an ellipsoidal shell can be described by
%%%%%%%%%%%%%%%%%%%%%%%%%%%%%%%%%%%%%%%%%%%%%%%%%%%%%%%%%%%%%%%%%%%%%%%%
\begin{equation}
\left\{ 
\begin{array}{l}
\displaystyle X = R \, \left(\frac{a}{c}\right) \, \sin \Theta \cos \Phi \\
\displaystyle Y = R \, \left(\frac{b}{c}\right) \, \sin \Theta \sin \Phi \\
Z = R \, \cos \Theta
\end{array}
 \right . ,
\label{eq:tricoord}
\end{equation}
%%%%%%%%%%%%%%%%%%%%%%%%%%%%%%%%%%%%%%%%%%%%%%%%%%%%%%%%%%%%%%%%%%%%%%%%
where $X$, $Y$ and $Z$ axes are the principal vectors of the
ellipsoidal shell, and $a/c$ and $b/c$ are the axis ratios.  In the
rest of the paper, we preferentially use the capital $R$ to refer to
the length of the major axis defined in the triaxial model.

Then the quadrupole moments of the iso-density surfaces in the
spherical and triaxial models, $Q_s$ and $Q_e$, are computed as
%%%%%%%%%%%%%%%%%%%%%%%%%%%%%%%%%%%%%%%%%%%%%%%%%%%%%%%%%%
\begin{eqnarray}
Q_s &\equiv& {1\over 5N_p}\sum_{m=-2}^{+2}
\left|\sum_{j} Y_{2m}(\theta_j,\phi_j)\right|^2-1  , \\
Q_e &\equiv& {1\over 5N_p}\sum_{m=-2}^{+2}
\left|\sum_j Y_{2m}(\Theta_j,\Phi_j)\right|^2 -1, 
\end{eqnarray}
%%%%%%%%%%%%%%%%%%%%%%%%%%%%%%%%%%%%%%%%%%%%%%%%%%%%%%%%%%%
where the summation over $j$ runs for all particles ($N_p$) in the
iso-density surface, and $Y_{lm}$ is the spherical harmonics. If the
spherical (triaxial) model is exact, $Q_s$ $(Q_e)$ vanishes.  Using
these measures, we will show the extent to which the triaxial model
indeed provides a significantly improved description for the simulated
halos.

In practice, we compute $Q_s(r)$ and $Q_e(R)$ for 5 shells of each
halo at $r=R=0.65r_{\rm vir}$, $0.35r_{\rm vir}$, $0.2r_{\rm
vir}$,$0.12r_{\rm vir}$, and $0.065r_{\rm vir}$ with the shell
thickness $\Delta r/r = \Delta R/R = \ln 10 \times 0.1=0.23$.  Those
shells are centered at the potential minimum of the halo. In the
triaxial model, we assume that { \it the shells have the same axis
ratios and the same principal axis directions as measured from the
isodensity surface at $A^{(3)}=2500$}.  Thus those shells do not
necessarily correspond to the iso-density surfaces that we have
discussed. Actually this treatment is important because otherwise the
triaxial model (with more degrees of freedom) should always provide a
better fit. Also this approximation is most likely what one would like
to apply statistically to halos of visible objects, which would yield
a practical and fair comparison between the spherical and triaxial
models.

\begin{figure}[htb]
\begin{center}
 \leavevmode\epsfxsize=8.0cm \epsfbox{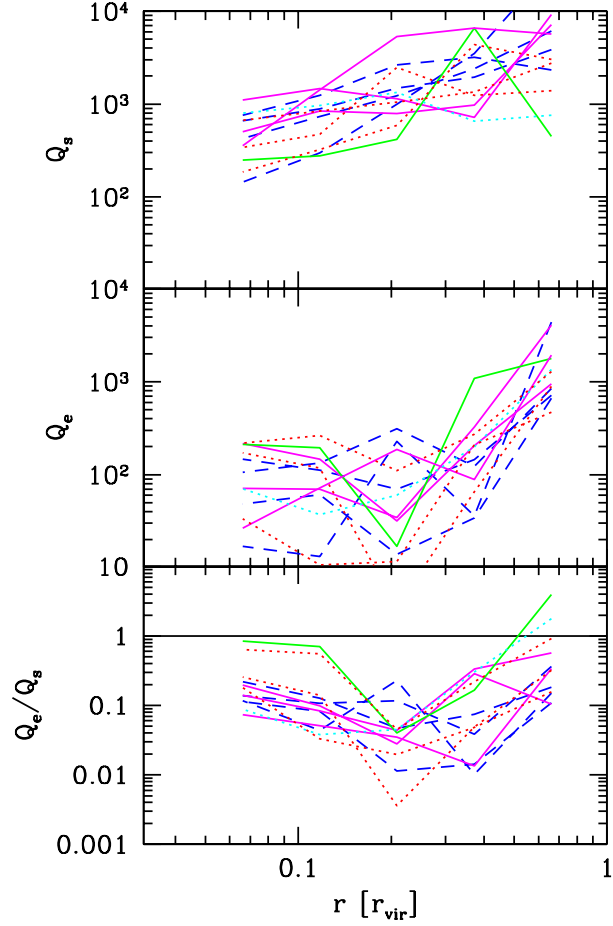}
\end{center} 
\figcaption{The quadrupole moments defined in the triaxial model
 ($Q_e$) and in the spherical model ($Q_s$) for five shells at radii
 from $0.05r_{\rm vir}$ to $0.65r_{\rm vir}$. They are presented in
 the top two panels, and their ratio $Q_e/Q_s$ is in the bottom
 panel. The dashed lines are for cluster halos, the dotted ones for
 group halos, and the solid lines for galactic halos. In the
 electronic edition, the cyan dotted line is for GR5 and the green
 solid one for GX3.  \label{fig:qeqs} }
\end{figure}

In the top and the middle panels of Figure \ref{fig:qeqs}, we present
the quadruple moments $Q_s$ and $Q_e$ for the twelve halos. The
quadruple $Q_s$ for the spherical modeling increases nearly
monotonically with the radius. The $Q_e$ for our ellipsoidal modeling
(in its simplified version as described above) stays flat at $R<0.3
r_{\rm vir}$ but increases with the radius at the larger radius. The
ratio of the two quadruples is shown in the bottom panel of Figure
\ref{fig:qeqs} which indicates that our triaxial model, even
simplified, fits the simulated halo profiles much better than the
spherical model. For 10 out of 12 halos, the ratio, $Q_e(R)/Q_s(r)$,
is much smaller than 1 at all scales ($r=R$).  Even for the remaining
two halos (GX3 and GR5), the ratio exceeds unity a bit only at the
largest radius, and the triaxial description shows a significant
improvement over the spherical model.  The ratio $Q_e(R)/Q_s(r)$ seems
to approach unity as $r$ becomes closer to $r_{\rm vir}$.  The reason
might be that the subclustering is more prominent in the outskirt
region than in the central region, since the subhalos are tidally
stripped when they fall into the central region (see Fig.~1 of
\citet{jingsuto00}). The strong subclustering makes both the spherical
model and the ellipsoidal model difficult to accurately describe the
complicated density distribution at $r_{vir}$. But the figure also
clearly shows that our triaxial model works significantly better than
the spherical model for $r<0.7 r_{vir}\approx r_{200}$, i.e. almost
the entire halo (the definition for $r_{200}$ will be given shortly).

\subsection{Density profiles in the triaxial model}

The next important task is to describe the density profiles in the
triaxial model generalizing the previous results in the spherical
approximation \citep[NFW; ][]{moore98,jingsuto00,k2001}.  In the same spirit
of the previous subsection, we do not perform the fit to the
iso-density surfaces that we identified, but rather compute the mean
density $\rho(R)$ at the simplified triaxial shells (i.e., the same
axis ratios and the axis directions for the entire halo as those
measured from its isodensity surface at $A^{(3)}=2500$) within a
thickness of $\Delta R/R = 0.12$.

\begin{figure}[htb]
\begin{center}
 \leavevmode\epsfxsize=15.0cm \epsfbox{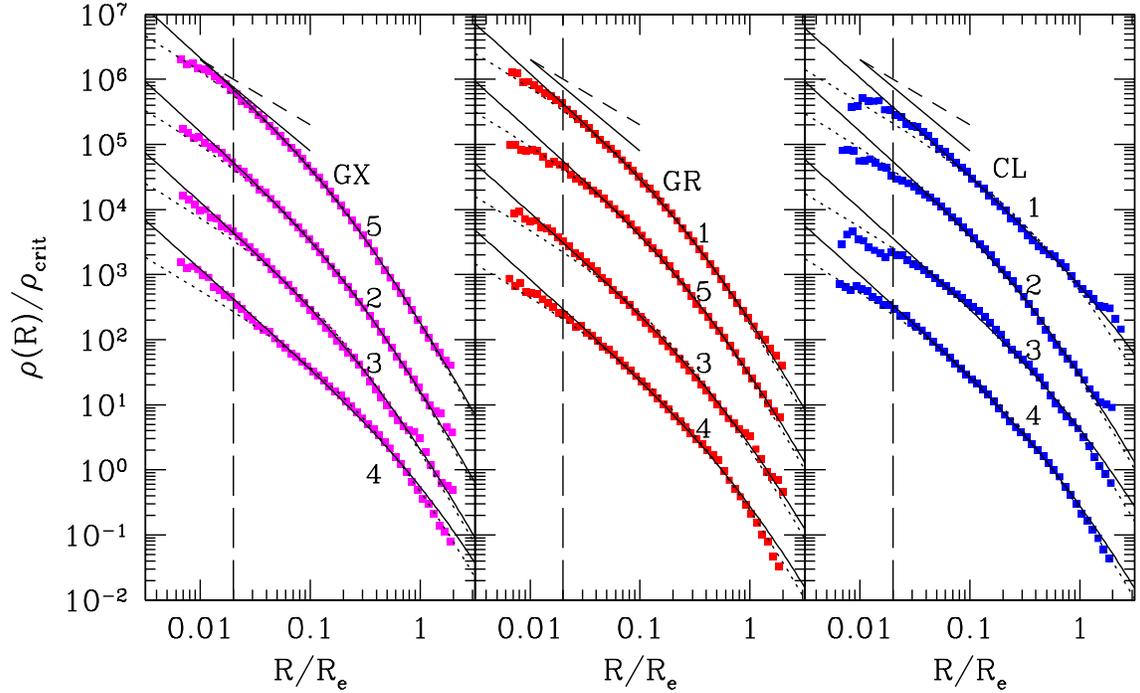}
\end{center} 
\figcaption{Radial density profiles in our triaxial model of the
  simulated halos of galaxy ({\it left}), group ({\it middle}), and
  cluster ({\it right}) masses.  The solid and dotted curves represent
  fits to equation (\ref{eq:nfw}) with $\alpha=1.5$ and 1.0,
  respectively.  For reference, we also show $\rho(R) \propto R^{-1}$
  and $R^{-1.5}$ in dashed and solid lines.  The vertical dashed lines
  indicate the force softening length which corresponds to our
  resolution limit.  For the illustrative purpose, the values of the
  halo densities are multiplied by 1, $10^{-1}$, $10^{-2}$, $10^{-3}$
  from top to bottom in each panel. 
 \label{fig:dens_prof} }
\end{figure}

Figure \ref{fig:dens_prof} plots the density profiles measured in this
way for individual halos as a function of $R$.  As in the spherical
case, we adopt the following form:
%%%%%%%%%%%%%%%%%%%%%%%%%%%%%%%%%%%%%%%%%%%%%%%%%%%%%%%%%%%%%%%
\begin{equation}
\label{eq:nfw}
\frac{\rho(R)}{\rho_{\rm crit}}=\frac{\delta_{\rm c}}
{\left(R/R_{\rm 0}\right)^\alpha\left(1+R/R_{\rm 0}\right)^{3-\alpha}},
\end{equation}
%%%%%%%%%%%%%%%%%%%%%%%%%%%%%%%%%%%%%%%%%%%%%%%%%%%%%%%%%%%%%%%
where $R_{\rm 0}$ is a scale radius and $\delta_{\rm c}$ is a
characteristic density.  Again following the definition of $r_{200}$
in the spherical model (within which the mean matter density is
$200\rho_{\rm crit}$), we define a radius $R_e$ so that the mean
matter density within the ellipsoid of the major axis radius $R_e$ is
$\Delta_e\rho_{\rm crit}$ with
%%%%%%%%%%%%%%%%%%%%%%%%%%%%%%%%%%%%%%%%%%%%%%%%%%%%%%%%%%%%%%%
\begin{equation}
\label{eq:deltae}
\Delta_e=5\Delta_{\rm vir} \left(\frac{c^2}{ab}\right)^{0.75}.
\end{equation}
%%%%%%%%%%%%%%%%%%%%%%%%%%%%%%%%%%%%%%%%%%%%%%%%%%%%%%%%%%%%%%%
The non-trivial dependence of $\Delta_e$ on the axis ratios in the
above equation is chosen so that $R_e$ becomes a fixed fraction of the
virial radius $r_{\rm vir}$ (see Fig.\ref{fig:concen} below).

The best-fits to equation (\ref{eq:nfw}) for each halo are shown in
Figure \ref{fig:dens_prof} for $\alpha=1.5$ (solid lines) and for
$\alpha=1.0$ (dotted lines).  Up to the resolution limit of the halo
simulations ($R/R_e\approx 0.02$), equation (\ref{eq:nfw}) yields a
good fit both for $\alpha=1$ and for $\alpha=1.5$.  If comparing the
fits to the simulation data more carefully, however, $\alpha=1$ works
better for the halos of cluster mass and $\alpha=1.5$ better for those
of galactic mass, which is consistent with the finding of Jing \& Suto
(2000) in the spherical model \citep[but see ][ for a different point
of view] {fukushige01}.

We also introduce a concentration parameter in our triaxial model:
%%%%%%%%%%%%%%%%%%%%%%%%%%%%%%%%%%%%%%%%%%%%%%%%%%%%%%%%%%%%%%%
\begin{equation}
\label{eq:ce}
c_{\rm e}\equiv \frac{R_{\rm e}}{R_0} \,,
\end{equation}
%%%%%%%%%%%%%%%%%%%%%%%%%%%%%%%%%%%%%%%%%%%%%%%%%%%%%%%%%%%%%%%
which is plotted in the upper panel of Figure \ref{fig:concen}
adopting $\alpha=1.0$ (crosses) and $\alpha=1.5$ (filled circles) in
the fit.  In what follows we will not address the issue related to the
inner slope of the density profiles, and adopt $\alpha=1$. It should
be noted, however, that our statistical results presented in the next
section can be readily applied to the $\alpha=1.5$ case since the
ratio $c_e(\alpha=1.5)/c_e(\alpha=1)$ is always close to 1/2.

Before moving to the statistical analysis of halos in the cosmological
simulations, we note that the value of $R_e$ and thus that of $c_e$ are
dependent on our specific definition of $\Delta_e$
(eq.[\ref{eq:deltae}]).  As the middle and bottom panels in
Figure~\ref{fig:concen} indicate, both $R_e/r_{\rm vir}$ and $c_e/c_{\rm
vir}$ (where $c_{\rm vir}$ is the ratio of the virial halo radius to the
scale radius $r_s$ in the spherical model) remain constant ($\approx
0.45$) independently of the mass of the halos when we adopt equation
(\ref{eq:deltae}) for $\Delta_e$.  This property is quite useful in
applying our results for a variety of theoretical predictions, since for
a halo of given virial mass $M_{vir}$:
%%%%%%%%%%%%%%%%%%%%%%%%%%%%%%%%%%%%%%%%%%%%%%%%%%%%
\begin{equation}
\label{eq:mvir}
M_{vir}=\frac{4\pi}{3} r^3_{vir} \Delta_{\rm vir} \rho_{\rm crit}\,,
\end{equation}
%%%%%%%%%%%%%%%%%%%%%%%%%%%%%%%%%%%%%%%%%%%%%%%%%%%%
the radius $R_e$ in our triaxial model is easily computed.  It is also
known that the $c_{\rm vir}$ is a function of the halo mass (NFW; Eke,
Navarro \& Steinmetz 2001) with the scatter described by the lognormal
distribution function (Jing 2000; Bullock et al. 2001).  Therefore,
once the shape of a halo at a given mass is specified, the density
profile of the halo is completely fixed.  The statistical distribution
function of the halo shape is discussed in the next section.

%%%%%%%%%%%%%%%%%%%%%%%%%%%%%%%%%%%%%%%%%%%%%%%%%%%%%%%%%%%
\section{Statistics of triaxial density profiles}
%%%%%%%%%%%%%%%%%%%%%%%%%%%%%%%%%%%%%%%%%%%%%%%%%%%%%%%%%%%

High-resolution halo simulations, like those used in the last section,
is well suited for studying the detailed internal structures of
individual halos, but the number of such halos is too small for a
statistical description. Therefore we switch to the halo catalogs
constructed from our cosmological simulations in order to study the
probability distribution of the shape of halos.  As emphasized in \S 2,
the cosmological simulations employ $N=512^3$ particles in a
100$h^{-1}$Mpc box and thus the mass resolution is even better than that
of {\it individual halo simulations} of the original NFW paper, for
instance.

We consider halos which contain more than $10^4$ particles within the
virial radius. The lower mass limits are $6.2\times 10^{12}\msun$ and
$2\times 10^{13}\msun$ in the LCDM and SCDM models, respectively. We
also consider three epochs at redshifts $z=0$, 0.5 and 1.0 to examine
the time-dependence. At these redshifts, we have 2494, 2160, and 1534
halos in the LCDM model, and 1806, 879, and 263 halos in the SCDM model,
respectively.

\subsection{Probability distribution of axis ratios}

Following the prescription presented in the last section, we determine
the halo shapes at the iso-density surfaces with $A^{(3)}=2500$.  Since
the typical radius of the surfaces is about $0.3r_{\rm vir}$, they are
well resolved in our cosmological simulations; the force softening
length is typically smaller by one order of magnitude.

Left panels of Figures \ref{fig:pac_lcdm} and \ref{fig:pac_scdm} present
the ratio $a/c$ of the minor axis $a$ to the major axis $c$ for halos
from the cosmological simulations in the LCDM and SCDM models,
respectively; solid, dotted and dashed histograms indicate the results
for $10^4 \le N_{\rm halo} < 2\times10^4$, $2\times10^4 \le N_{\rm halo}
< 6\times10^4$, $6\times 10^4 \le N_{\rm halo}$, where $N_{\rm halo}$ is
the number of particles within the virial radius of each halo; in those
Figures we use $M_4 \equiv N_{\rm halo}/10^4$, and thus $M_4=1$
corresponds to $M_{\rm vir} = 6.2\times10^{12}h^{-1}M_\odot$, and
$2.1\times10^{13}h^{-1}M_\odot$ for our LCDM and SCDM models). Top,
middle, and bottom panels show the results at $z=0$, 0.5, and 1.0.

Two systematic trends are visible; the ratio is slightly larger for less
massive halos, and decreases at higher redshifts. 
This motivates us to attempt the following empirical scaling for the
axis ratio $a/c$:
%%%%%%%%%%%%%%%%%%%%%%%%%%%%%%%%%%%%%%%%%%%%%%%%%%%%%%%%%%%%%%%%%%%%%%
\begin{equation}
\label{eq:acs}
\tilde{r}_{\rm ac} \equiv \left(\frac{a}{c}\right)_{\rm sc}
= \left(\frac{a}{c}\right) 
\left(\frac{M_{\rm vir}}{M_{\star}}\right)^{0.07[\Omega(z)]^{0.7}}\,,
\end{equation}
%%%%%%%%%%%%%%%%%%%%%%%%%%%%%%%%%%%%%%%%%%%%%%%%%%%%%%%%%%%%%%%%%%%%%%
where $M_{\star}$ is the characteristic non-linear mass at $z$ so that
the rms top-hat smoothed overdensity at the scale
$\sigma(M_{\star},z)$ is $\delta_{\rm c} = 1.68$.  The $M_{\star}$ at
$z=0$, 0.5, and 1.0 are $9.4\times 10^{12}h^{-1}M_\odot$, $2.0\times
10^{12}h^{-1}M_\odot$, and $3.8\times 10^{11}h^{-1}M_\odot$
respectively for LCDM, and $8.5\times 10^{12}h^{-1}M_\odot$,
$9.7\times 10^{11}h^{-1}M_\odot$, and $1.4\times 10^{11}h^{-1}M_\odot$
respectively for SCDM.

Such scaled axis ratios $\tilde{r}_{\rm ac}$
show a fairly universal distribution almost independently of the mass
and the epoch (histograms in the the right panels of
Figs. \ref{fig:pac_lcdm} and \ref{fig:pac_scdm}). The {\it universal}
probability distribution function of the ratio $\tilde{r}_{\rm ac}$ is
well fitted to the following Gaussian:
%%%%%%%%%%%%%%%%%%%%%%%%%%%%%%%%%%%%%%%%%%%%%%%%%%%%%%%%%%%%%%%%%%%%%
\begin{equation}
\label{eq:pacfit}
p(\tilde{r}_{\rm ac})\, d\tilde{r}_{\rm ac}
 = \frac{1}{\sqrt{2\pi} \sigma_s}
\exp \left({-\frac{(\tilde{r}_{\rm ac}-0.54)^2}{2\sigma^2_s}}\right)
\, d\tilde{r}_{\rm ac}
\end{equation}
%%%%%%%%%%%%%%%%%%%%%%%%%%%%%%%%%%%%%%%%%%%%%%%%%%%%%%%%%%%%%%%%%%%%%
with $\sigma_s=0.113$.

Next we decompose the joint probability distribution function of the
axis ratios as
%%%%%%%%%%%%%%%%%%%%%%%%%%%%%%%%%%%%%%%%%%%%%%%%%%%%%%%%%%%%%%%%
\begin{eqnarray} 
\label{eq:pcon} 
p(a/c,b/c)d(a/c)d(b/c) &=& p(a/c)d(a/c) ~ p(b/c|a/c)d(b/c) \cr
&=& p(a/c)d(a/c) ~ p(a/b|a/c)d(a/b)
\end{eqnarray} 
%%%%%%%%%%%%%%%%%%%%%%%%%%%%%%%%%%%%%%%%%%%%%%%%%%%%%%%%%%%%%%%%%
in terms of the conditional probability distribution functions,
$p(b/c|a/c)$ and $p(a/b|a/c)$.  The second equality holds because once
$a/c$ is fixed, the distribution of $a/b$ is uniquely determined from
that of $b/c$. Since we have shown that the distribution function
$p(a/c)$ is well approximated by equations (\ref{eq:acs}) and
(\ref{eq:pacfit}), we compute the conditional probability distribution
$p(a/b|a/c)$. Figures \ref{fig:pab_lcdm} and \ref{fig:pab_scdm} plot
the results for the LCDM and SCDM models, respectively. Different panels
correspond to $p(a/b|a/c)$ for different ranges of $a/c$. Solid, dotted,
and dashed histograms indicate $p(a/b|a/c)$ at $z=0$, 0.5, and 1.0,
respectively.

The conditional functions appear to be insensitive to the redshift.  In
both cosmological models, they are accurately fitted to
%%%%%%%%%%%%%%%%%%%%%%%%%%%%%%%%%%%%%%%%%%%%%%%%%%%%%%%%%%%%%%
\begin{equation}
\label{eq:pabfit} 
p(a/b|a/c)= \frac{3}{2(1-r_{\rm min})}
\left[1-\left(\frac{2a/b-1-r_{\rm min}}{1-r_{\rm min}}\right)^2\right] ,
\end{equation}
%%%%%%%%%%%%%%%%%%%%%%%%%%%%%%%%%%%%%%%%%%%%%%%%%%%%%%%%%%%%%%%%
for $a/b\ge r_{\rm min}$, where $r_{\rm min}=a/c$ for $a/c\ge 0.5$ and
$r_{\rm min}=0.5$ for $a/c<0.5$.  $p(a/b|a/c)=0$ for $a/b\le r_{\rm
min}$.

\subsection{Probability distribution of the concentration parameter}

We apply the triaxial density profile (eq.[\ref{eq:nfw}]) obtained in
the halo simulations to the halo catalogs in the cosmological
simulations.  Considering the resolution limits, we adopt $\alpha=1$ and
use the data points at $ \eta < R_e < r_{\rm vir}$ in the fit, where
$\eta$ the force softening length (see \S 2). Since we do not address
the innermost structures of the halos and rather focus on the value of
the concentration parameter $c_e$, this catalog has a sufficient
resolution to yield an unbiased estimate (e.g., Jing 2000; Bullock et
al. 2001; Eke, Navarro \& Steinmetz 2001 for discussion).  As already
found in the spherical model (Jing 2000), the distribution of $c_e$ in
the triaxial model has a significant scatter even if the range of the
halo mass is fairly specified reflecting the dependence of the merging
history of the individual halo.

The resulting probability distribution functions for $c_e$ are
presented in Figure \ref{fig:chist}, which are well fitted by the
lognormal distribution:
%%%%%%%%%%%%%%%%%%%%%%%%%%%%%%%%%%%%%%%%%%%%%%%%%%%%%%%%%%%%%%%%%%%%%%
\begin{equation}
\label{eq:lognormal}
p(c_e)dc_e={1\over \sqrt{2\pi}\sigma_{c_e}}
\exp \left[-{(\ln c_e-\ln \bar c_e)^2\over 2\sigma_{c_e}^2}\right]
\, d\ln c_e \, 
\end{equation}
%%%%%%%%%%%%%%%%%%%%%%%%%%%%%%%%%%%%%%%%%%%%%%%%%%%%%%%%%%%%%%%%%%%
with a dispersion of $\sigma_{c_e} \approx 0.3$ both in the SCDM and
LCDM models.  The dispersion is slightly larger than the value
estimated in the spherical model ($\approx 0.2$) for equilibrium
halos, but is comparable to the value for all halos put together (Jing
2000).  It should be noted here that despite the fact that the
triaxial model is superior in describing the density distribution of
halos, the scatter in $c_e$ is comparable to the scatter of
concentrations in spherical profile fits, which probably means that
the scatter originates from the halo merger histories rather than
non-sphericity of the halos.

The probability distribution (eq.[\ref{eq:lognormal}]) is completed by
specifying the mean of the concentration parameter $\bar c_e$. The
result from our simulations is plotted in Figure \ref{fig:cfit} as a
function of the halo mass at $z=0$, 0.5 and 1.0.  NFW proposed a
semi-analytic fitting formula for the concentration $c_{\rm vir}$ in
the spherical model.  \footnote{Originally NFW defined the
concentration parameter as $c_{200} \equiv r_{200}/r_s$, where
$r_{200}$ is the radius within which the mean overdensity is
$200\rho_{\rm crit}$.  Their recipe, however, can be easily
generalized to $c_{\rm vir}$, since $r_{200}/r_{v}$ is almost constant
for a given cosmology.}  More recently Bullock et al. (2001) have
shown that in their LCDM model (the parameters are similar to those of
our LCDM model here) $c_{\rm vir}$ of a given mass decreases with $z$
as $\propto (1+z)^{-1}$.  The redshift dependence is stronger than
that predicted in the NFW recipe.  Thus Bullock et al. (2001) have
proposed another recipe which successfully describes the concentration
$c_{\rm vir}$.  Since we have already shown that the ratio,
$c_e/c_{\rm vir}$, is almost constant (Fig.\ref{fig:concen}), it is
interesting to see if the formula of Bullock et al. (2001) also
describes the behavior of $c_e$ in our triaxial model.

In the LCDM model, we find that the redshift dependence of $c_e$ for a
given mass is approximately $\propto (1+z)^{-1}$ in good agreement
with their result. In the SCDM model, however, our result of $c_e$
shows a stronger redshift dependence than their prediction.  This is
also likely that the fitting formula of Bullock et al. (2001) was
designed to fit spherical concentrations, $c_{\rm vir}$, and is not
applicable to the non-spherical case due to the evolution of
$c_e/c_{\rm vir}$.

 Following NFW and Bullock et al. (2001), we propose a new fitting
formula for $\bar c_e$ in the triaxial model:
%%%%%%%%%%%%%%%%%%%%%%%%%%%%%%%%%%%%%%%%%%%%%%%%%%%%%%%%%%%%%%%%%%%%
\begin{equation}
\label{eq:cefitting}
\bar c_e(M,z)=A_e 
\sqrt{\frac{\Omega(z)}{\Omega(z_c)}}
\left(\frac{1+z_c}{1+z}\right)^{\frac{3}{2}}.
\end{equation}
%%%%%%%%%%%%%%%%%%%%%%%%%%%%%%%%%%%%%%%%%%%%%%%%%%%%%%%%%%%%%%%%%%%
In the above, $z_c$ is the {\it collapse} redshift of the halo of mass $M$
(NFW):
%%%%%%%%%%%%%%%%%%%%%%%%%%%%%%%%%%%%%%%%%%%%%%%%%%%%%%%%%%%%%%%%%%%%%
\begin{equation}
\label{eq:zcps}
{\rm erfc}{\frac{\delta_c(z_c)-\delta_c}
{\sqrt{2[\sigma^2(fM)-\sigma^2(M)]}}}
=\frac{1}{2} ,
\end{equation} 
%%%%%%%%%%%%%%%%%%%%%%%%%%%%%%%%%%%%%%%%%%%%%%%%%%%%%%%%%%%%%%%%%%%%%%
where $\sigma(M)$ is the rms top-hat mass variance at $z=0$,
$\delta_c=1.68$, $\delta_c(z)=1.68/D(z)$, $D(z)$ is the linear growth
factor, and $f=0.01$.  Solid lines in Figure \ref{fig:cfit} indicate
the predictions of equation (\ref{eq:cefitting}), implying that the
formula describes our simulation results very accurately.  In those
plots, we adopt $A_e=1.1$ and 1.0 for the LCDM and SCDM models,
respectively.

We also made sure that the formula also agrees well with our halo
simulations in the LCDM model, while the results appear 10 to 20 \%
higher (i.e., $A_e=1.2\sim1.3$) than those of the cosmological
simulations ($A_e=1.1$). Considering both the typical 30\% scatter in
$c_e$ and the limited number of the high-resolution halos (12 in total),
the above level of difference may not be interpreted so seriously at
this point.  In fact, the difference may be attributed partly to the
fact that halos with significant substructures (like GR2) have been
eliminated in the high-resolution halo samples (\S 2), while we have not
attempted such a selection in the cosmological simulations. Indeed Jing
(2000) has noted that halos in equilibrium are systematically more
centrally concentrated than those with significant substructures.  We
also note that most previous studies including NFW and Eke, Navarro \&
Steinmetz (2001) have preferentially selected isolated halos in
re-simulating with higher resolution, which would have less
substructures and therefore have slightly higher concentration than
average.  If one is interested in halos in nearly equilibrium, the
best-fit value of $A_e$ should become 1.3. Since $c_e/c_{vir}$ remains
constant(Figs.\ref{fig:concen} and \ref{fig:cvsac}), the fitting formula
(eq.[\ref{eq:cefitting}]) can also be used for predicting $c_{vir}$ in
CDM models.

Finally we have checked if the fitted values of $R_e$ and $c_e$ are
dependent on the shapes of halos. Figure \ref{fig:radiusratio}
presents the ratio of $R_e$ to the virial radius $r_{\rm vir}$ as a
function of the axis ratio $a/b$. Clearly $R_e/r_{\rm vir}$ is
independent of $a/b$ and of the redshift (or equivalently the halo
mass in unit of $M_{\star}$, see also Fig.\ref{fig:concen}), and
approximately given by 0.45.  Similarly, we find that $R_e/r_{\rm
vir}$ is independent of $b/c$ and $a/c$. On the other hand, the
concentration parameter $c_e$ is slightly dependent on the halo
shape. Figure \ref{fig:cvsac} indicates that halos with smaller $a/c$
are less centrally concentrated.

In terms of the scaled axis ratio $(a/c)_{\rm sc}$ (eq.[\ref{eq:acs}]),
the ratio of the mean concentration $c_e$ for a given 
$\tilde{r}_{ac} = (a/c)_{\rm sc}$
and the overall average $\bar c_e(M,z)$ (eq.[\ref{eq:cefitting}]) is
well approximated by
%%%%%%%%%%%%%%%%%%%%%%%%%%%%%%%%%%%%%%%%%%%%%%%%%%%%%%%%%%%%%%%%%%%%
\begin{equation}
\label{eq:cvsacfit}
\frac{c_e[\tilde{r}_{ac},M,z]}{\bar c_e(M,z)}
=1.35\exp\left[-\left(\frac{0.3}{\tilde{r}_{ac}}\right)^2\right] .
\end{equation}
%%%%%%%%%%%%%%%%%%%%%%%%%%%%%%%%%%%%%%%%%%%%%%%%%%%%%%%%%%%%%%%%%%%%
This fit is plotted in the solid line in Figure \ref{fig:cvsac}, which
is in good agreement with the simulation data for different halo masses
and both in the LCDM and SCDM models.

In this section, we used the halos identified from cosmological
simulations which have $N>10^4$ particles. The smallest halos are
resolved much more poorly than the largest halos and the
high-resolution halos (\S 3) that consist of $\sim 10^6$ particles. In
order to make sure that our results are robust to the mass resolution, we
repeat the same analysis of the axis ratios and the density profile by
randomly selecting $N=10^4$ particles from each of the twelve
high-resolution halos of the previous section. The results are
compared in Figure \ref{fig:sim_res} with those obtained in the
last section where we consider all particles. Both the axis ratios and
the concentration of the randomly selected sample agree well with
those of the original halo sample; the typical dispersion of the axis
ratios between the two samples is $\sim 10\%$, and the concentration
$c_e$ of the randomly selected sample is slightly lower ($\sim
8\%$). This comparison indicates that the mass resolution does not
affect our results in this section significantly.

%%%%%%%%%%%%%%%%%%%%%%%%%%%%%%%%%%%%%%%%%%%%%%%%%%%%%%%%%%%%%%%
\section{Summary and Discussion}
%%%%%%%%%%%%%%%%%%%%%%%%%%%%%%%%%%%%%%%%%%%%%%%%%%%%%%%%%%%%%%%

This paper has presented a triaxial modeling of the dark matter halo
density profiles extensively on the basis of the combined analysis of
the high-resolution halo simulations (12 halos with $N\sim 10^6$
particles within their virial radius) and the large cosmological
simulations (5 realizations with $N=512^3$ particles in a $100h^{-1}$Mpc
boxsize). In particular, we found that the universal density profile
discovered by NFW in the spherical model can be also generalized to our
triaxial model description. Our triaxial density profile is specified by
the concentration parameter $c_e$ and the scaling radius $R_0$ (or the
{\it virial} radius $R_e$ in the triaxial modeling) as well as the axis
ratios $a/c$ and $a/b$.

\begin{figure}[htb]
\begin{center}
\vspace*{6cm}
 \leavevmode\epsfxsize=8.0cm \epsfbox{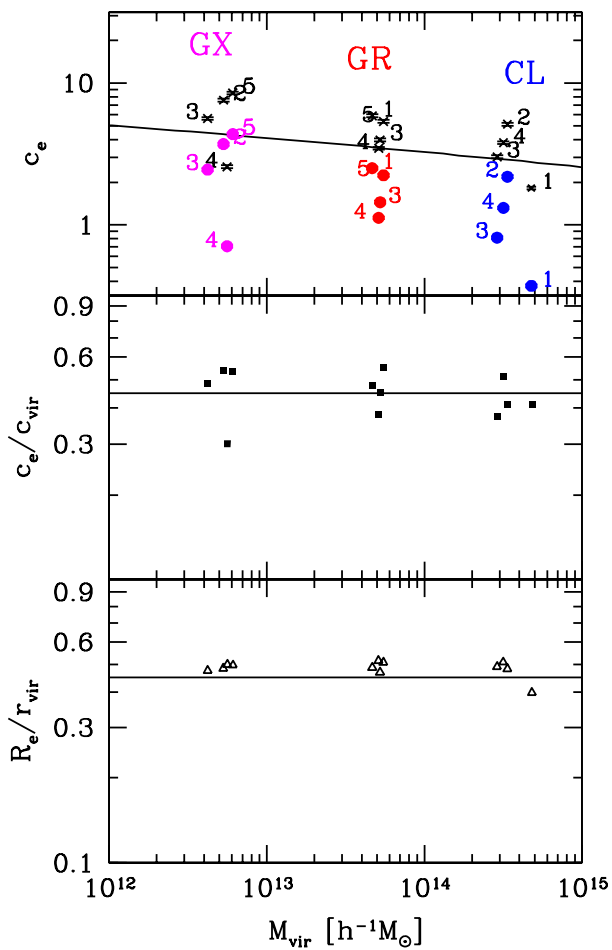}
\end{center} 
\figcaption{The fitting results of the triaxial model to twelve
 halos. a) the concentration parameter $c_e$ for $\alpha=1$ (crosses)
 and for $\alpha=1.5$ (filled circles); b) the ratio of $c_e$ to that of
 the spherical counterpart, $c_{\rm vir}$, for $\alpha=1$; c) the ratio
 of $R_e$ to the virial radius $r_{\rm vir}$ in the spherical model.
 \label{fig:concen} }
\end{figure}

\begin{figure}[htb]
\begin{center}
 \leavevmode\epsfxsize=12.0cm \epsfbox{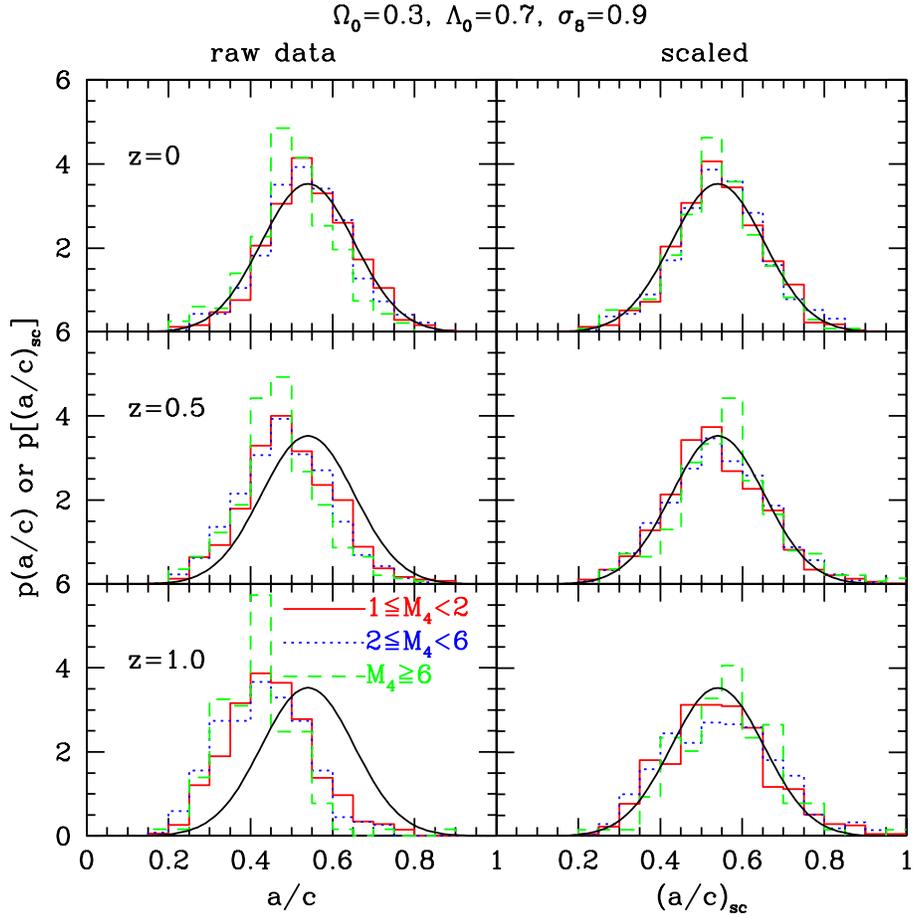}
\end{center} 
\figcaption{The distribution of the axis ratio $a/c$ of the halos in the
cosmological simulations of the LCDM model before ({\it left}) and after
({\it right}) the scaling described in the text. Top, middle and bottom
panels correspond to $z=0$, 0.5, and 1.0, respectively.  Solid, dotted
and dashed histograms indicate the results for halos that have the
number of particles of $M_4 \equiv (N_{\rm halo}/10^4)$ within the
virial radius.  The smooth solid curves in all the panels represent our
fit (eq.[{\ref{eq:pacfit}}]).  \label{fig:pac_lcdm} }
\end{figure}

\begin{figure}[htb]
\begin{center}
 \leavevmode\epsfxsize=12.0cm \epsfbox{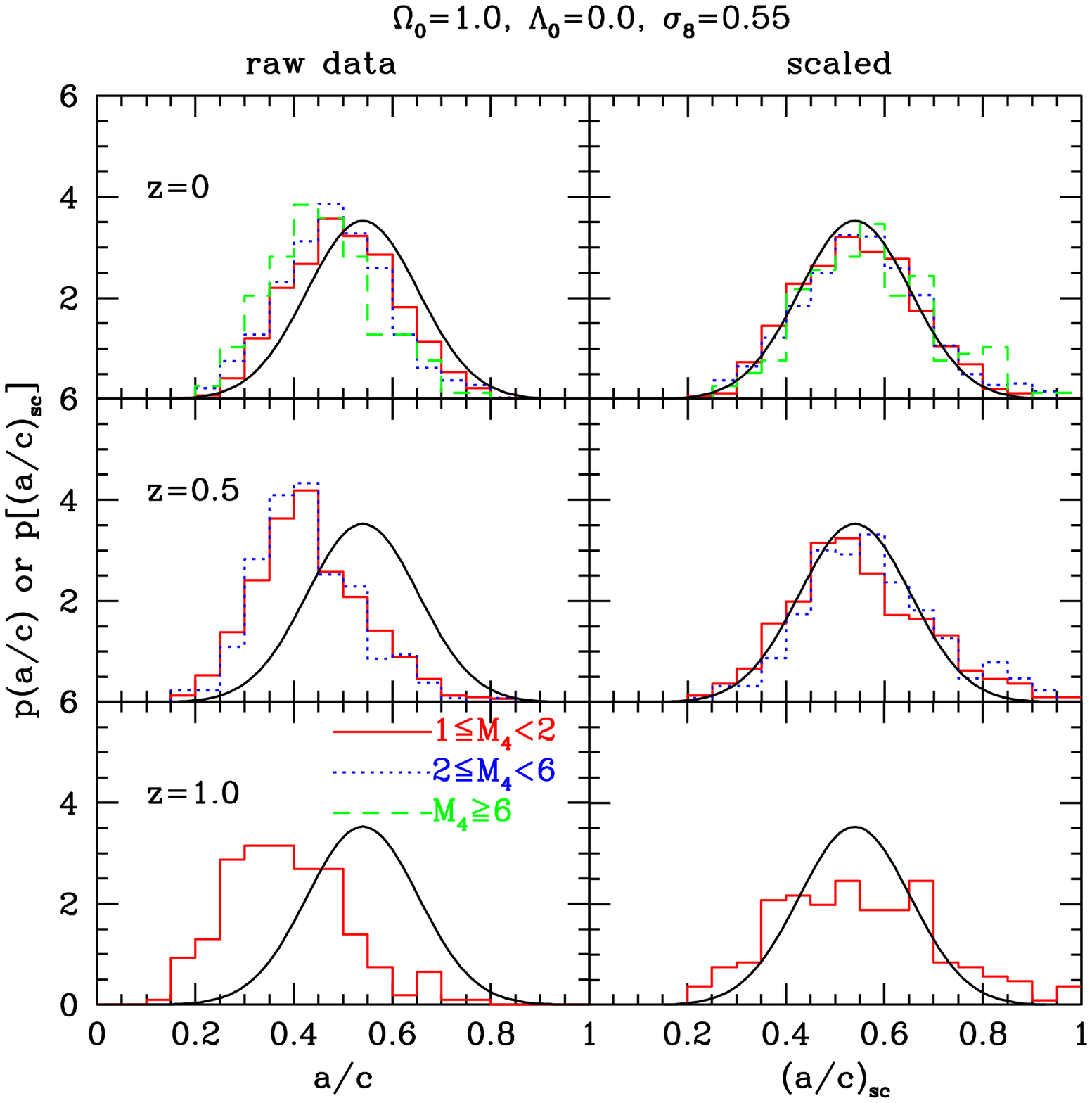}
\end{center} 
\figcaption{Same as Figure \ref{fig:pac_lcdm}, except for the halos in
the SCDM simulations.
\label{fig:pac_scdm} }
\end{figure}

\begin{figure}[htb]
\begin{center}
 \leavevmode\epsfxsize=12.0cm \epsfbox{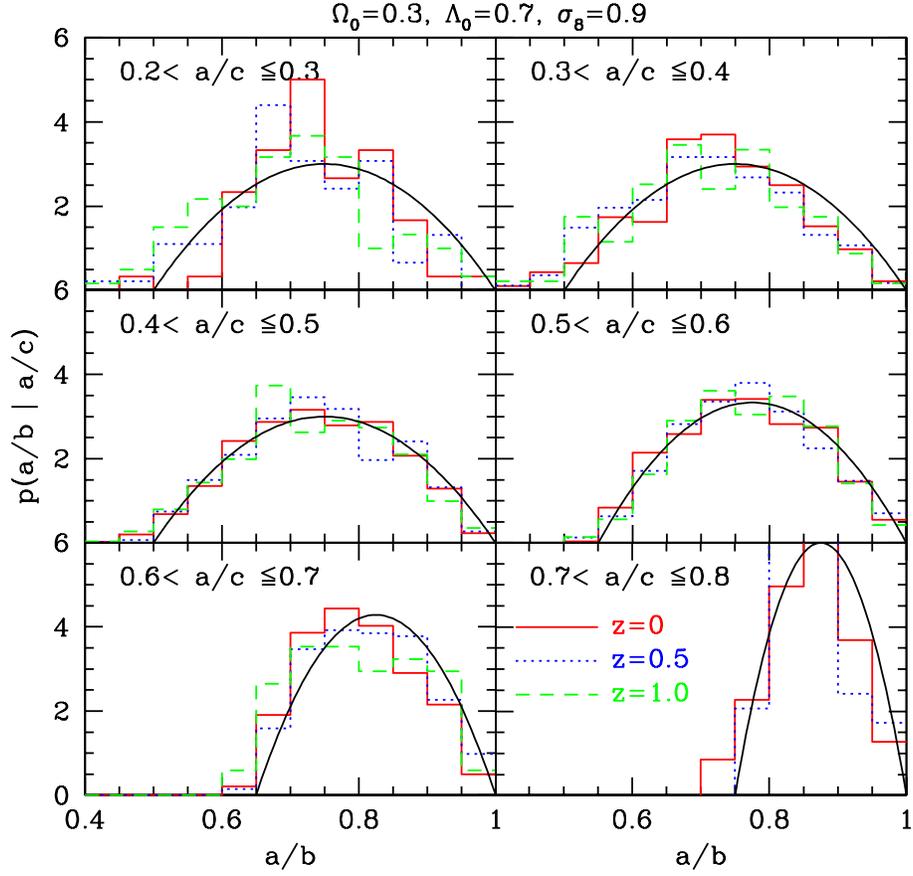}
\end{center} 
\figcaption{The conditional distribution of the axis ratio $a/b$ of the
halos in the cosmological simulations of the LCDM model for a given
range of $a/c$.  Halos at different redshifts are represented with
different lines as indicated in the bottom-right panel.  The smooth
solid curves in all the panels represent our fit
(eq.[{\ref{eq:pabfit}}]).  \label{fig:pab_lcdm} }
\end{figure}

\begin{figure}[htb]
\begin{center}
 \leavevmode\epsfxsize=12.0cm \epsfbox{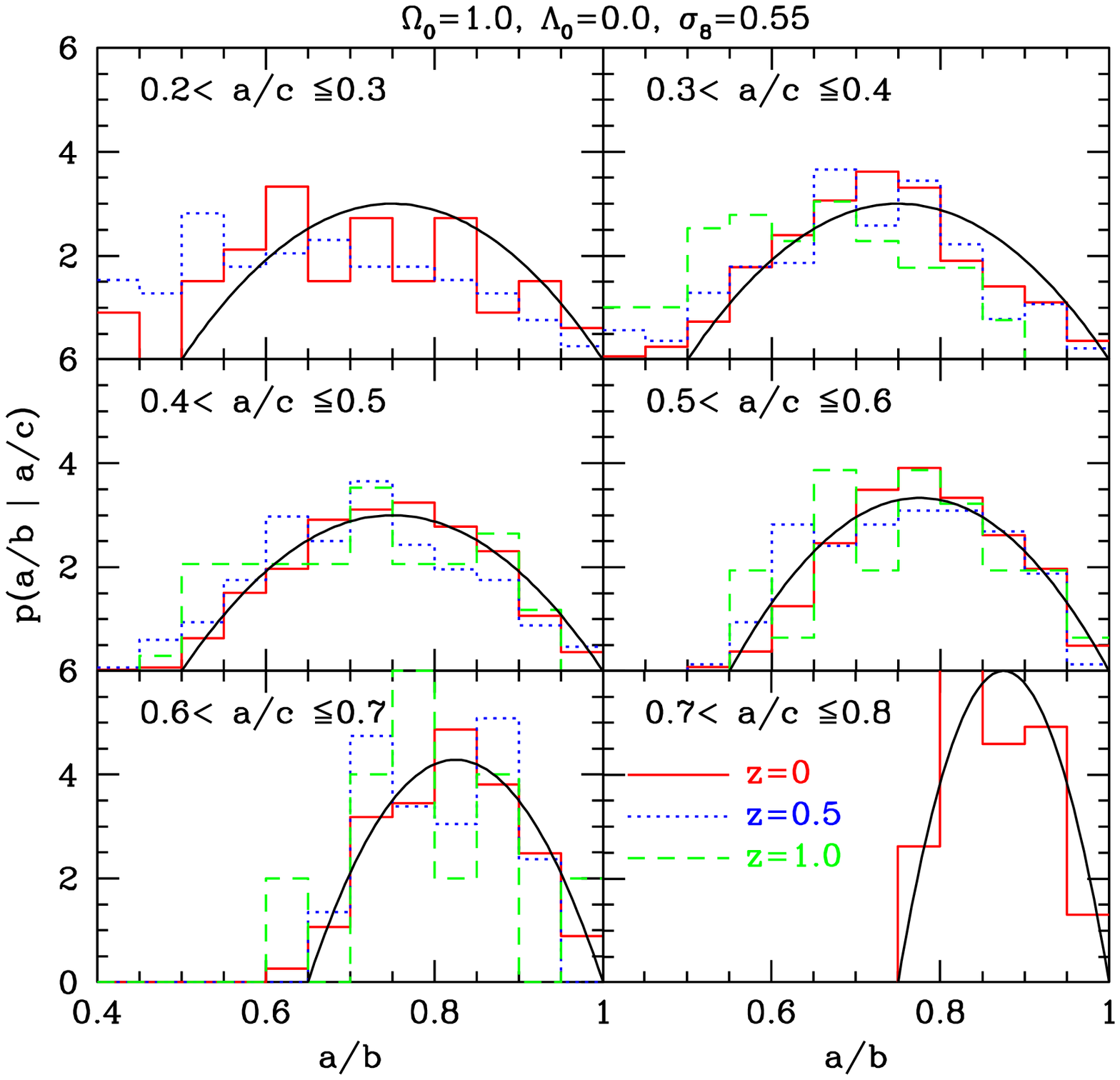}
\end{center} 
\figcaption{Same as Figure \ref{fig:pac_lcdm}, except for the halos in
the SCDM simulations.
\label{fig:pab_scdm} }
\end{figure}

\begin{figure}[htb]
\begin{center}
 \leavevmode\epsfxsize=12.0cm \epsfbox{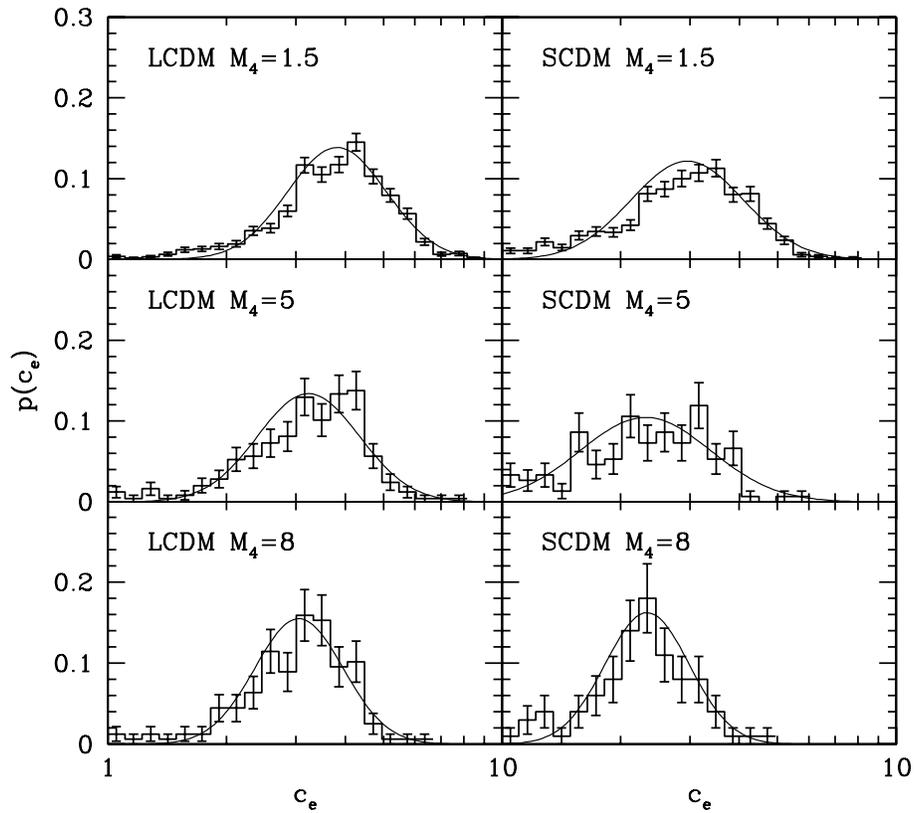}
\end{center} 
\figcaption{Distribution of the concentration $c_e$ of the halos in the
LCDM ({\it left}) and in the SCDM ({\it right}) models for different
halo mass $M_4 \equiv (N_{\rm halo}/10^4)$.  The smooth solid curves
represent our log-normal fit (eq.[\ref{eq:lognormal}]).
\label{fig:chist} }
\end{figure}

\begin{figure}[htb]
\begin{center}
 \leavevmode\epsfxsize=12.0cm \epsfbox{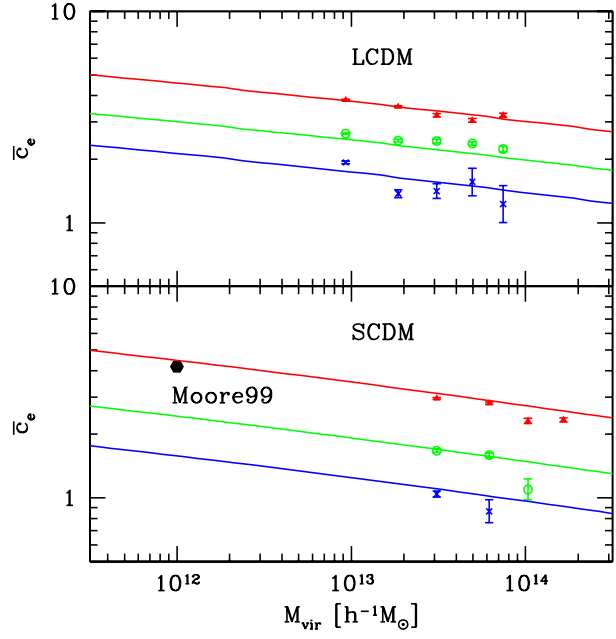}
\end{center} 
\figcaption{ The mean of the concentration $\bar c_e$ as a function of
the virial mass in the LCDM and SCDM models. The solid curves
represent our fitting formula (eq.[\ref{eq:cefitting}]) at $z=0$, 0.5,
and 1.0 from top to bottom. The data point, labeled Moore99, is taken
from the result of Moore et al. (1999), and is scaled according to our
fitting formula (eq.[\ref{eq:cefitting}]).
\label{fig:cfit} }
\end{figure}

\begin{figure}[htb]
\begin{center}
 \leavevmode\epsfxsize=12.0cm \epsfbox{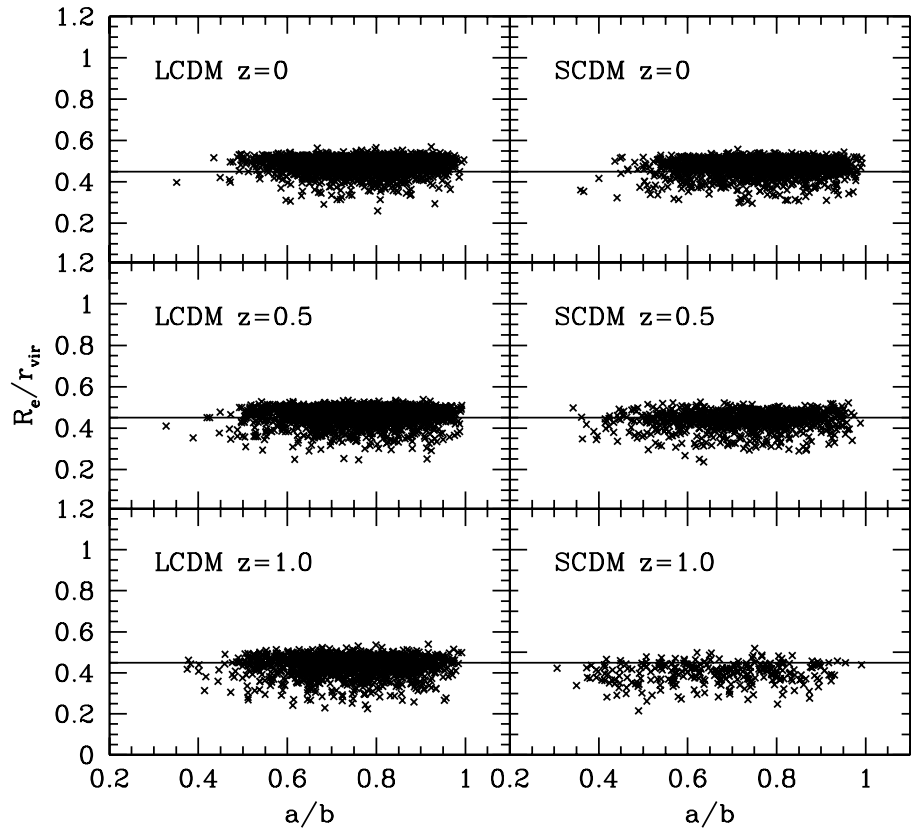}
\end{center} 
\figcaption{ The ratio of $R_e$ to the virial radius for halos with
different shapes in the LCDM ({\it left}) and SCDM ({\it right}) models
at different redshifts.  \label{fig:radiusratio} }
\end{figure}

\begin{figure}[htb]
\begin{center}
 \leavevmode\epsfxsize=12.0cm \epsfbox{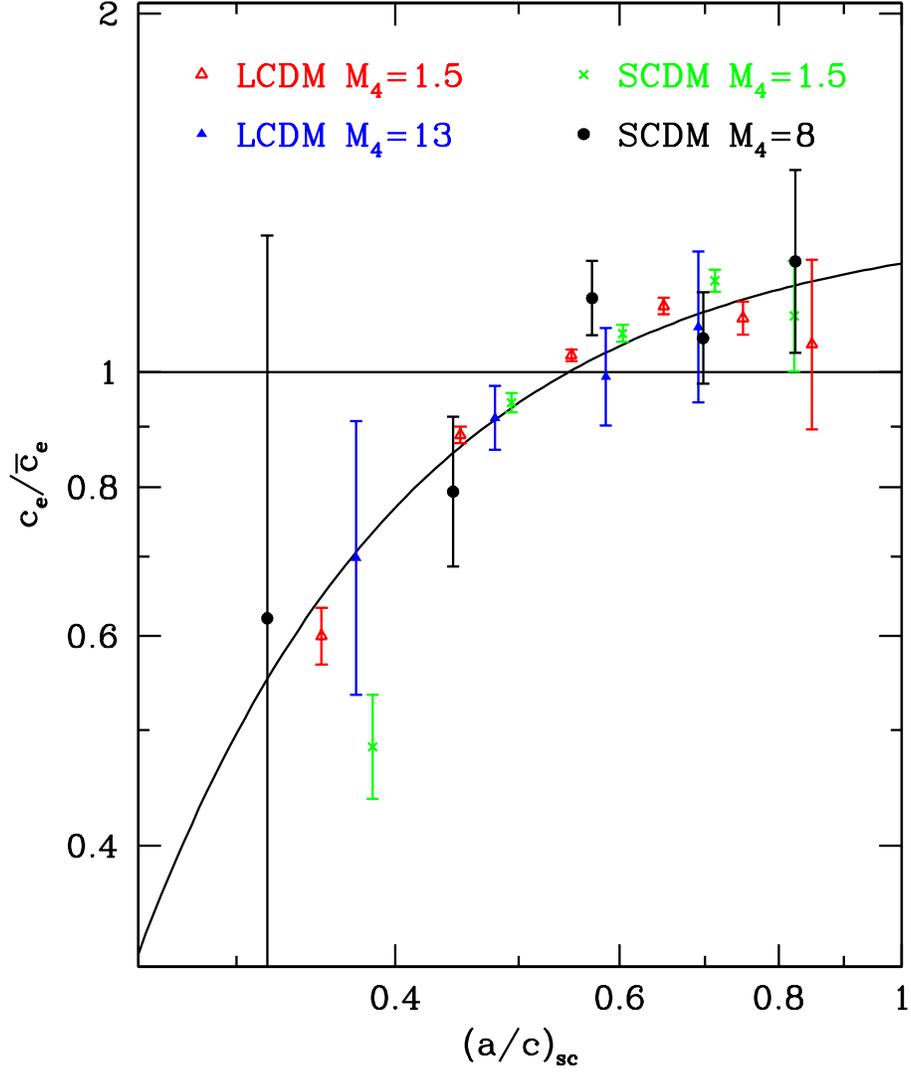}
\end{center} 
\figcaption{ The dependence of the ellipsoid concentration $c_e$ on the
 scaled axis ratio $(a/c)_{\rm sc}$.  Different symbols denote the
 results of halos of different mass ($N_{\rm halo} = 10^4M_4$ particles) 
in the LCDM and SCDM models.  
The smooth solid curve represents our fitting formula
 (eq.[\ref{eq:cvsacfit}]).  \label{fig:cvsac} }
\end{figure}

\begin{figure}[htb]
\begin{center}
 \leavevmode\epsfxsize=8.0cm \epsfbox{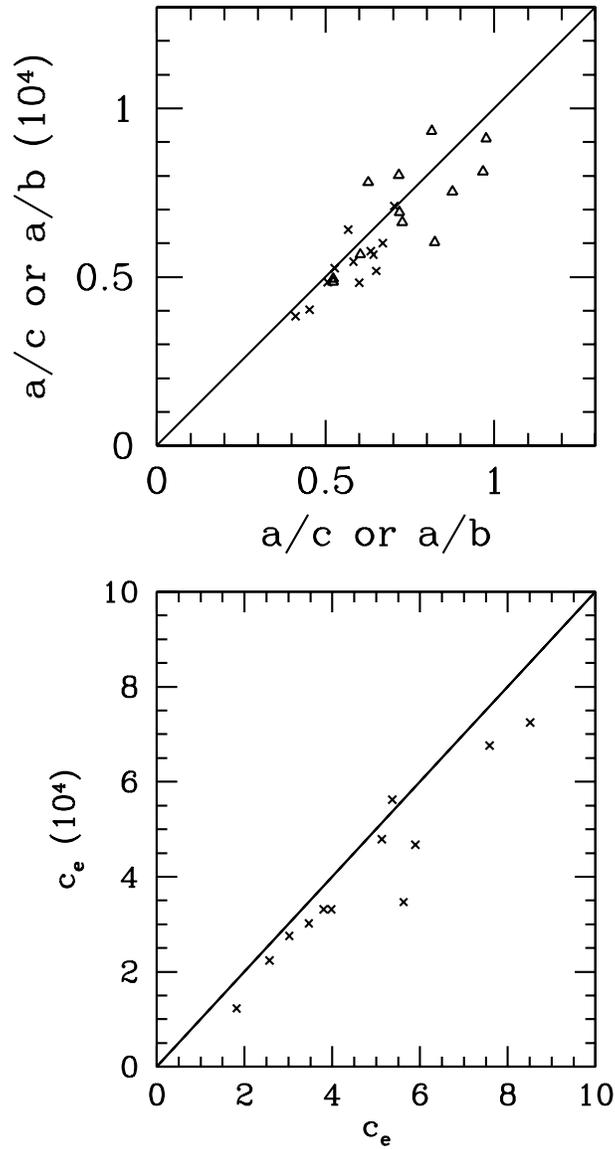}
\end{center} 
\figcaption{The axis ratios $a/c$ (crosses, upper panel) and $b/c$
(triangles, upper panel), and the concentration $c_e$ (lower panel) of
a subsample of halo particles randomly selected from the twelve
high-resolution halos compared with those of the whole sample
(abscissa). Each halo in the subsample has $N_p=10^4$ particles.
\label{fig:sim_res} }
\end{figure}

We have obtained several fitting formulae for those parameters which are
of practical importance in exploring the theoretical and observational
consequences of our triaxial model (in doing so we have adopted $\alpha=1$
since the precise value of the inner slope is difficult to reliably
determine even with the resolution of the current simulations);
%%%%%%%%%%%%%%%%%%%%%%%%%%%%%%%%%%%%%%%%%%%%%%%%%%%%%%%%%%%%%%%%%%%%%%%%%
\begin{itemize}
\item the mass and redshift dependence of the axis ratio, or
equivalently the definition of the scaled axis ratio $\tilde{r}_{ac}
\equiv (a/c)_{\rm sc}$ : equation(\ref{eq:acs})
\item the probability distribution of the axis ratio $p(\tilde{r}_{ac})$
: equation(\ref{eq:pacfit})
\item  the conditional probability distribution of the
axis ratios $p(a/b |a/c)$ : equation(\ref{eq:pabfit})
\item the mean value of the concentration parameter $\bar c_e(M,z)$ 
: equation(\ref{eq:cefitting})
\item the dependence of the concentration parameter on the axis ratio
$\tilde{r}_{ac}$ 
: equation(\ref{eq:cvsacfit})
\item the probability distribution of the concentration parameter 
$p(c_e)$ : equation(\ref{eq:lognormal})
\end{itemize}
%%%%%%%%%%%%%%%%%%%%%%%%%%%%%%%%%%%%%%%%%%%%%%%%%%%%%%%%%%%%%%%%%%%%%%%%%%
Since $c_e/c_{vir}$ remains constant(Figs.\ref{fig:concen} and
\ref{fig:cvsac}), the fitting formula (eq.[\ref{eq:cefitting}]) can
also be used for predicting $c_{vir}$ in CDM models.

We have focused on the triaxial modeling and characterization of dark
halos in the present paper, and plan to show specific applications
elsewhere. Nevertheless it would be worthwhile to mention several
important examples of the current model.

The results of the paper are applicable in a fairly direct
manner to the following three areas. (i) the weak and strong lens
statistics. The comparison with the weak lensing observations provides
information of the degree of triaxiality of observed clusters, mainly at
outer regions. In addition, the frequency of the lensing arc is known to
be sensitive to the non-sphericity of the halo mass profile especially
in the central regions
\citep[e.g.,][]{bart98,meneghetti00,meneghetti01,molikawa01,oguri02}.
(ii) predictions of the non-linear clustering of dark matter based on
the halo model \citep[e.g.,][]{MJB1997, MF2000, Hamanaetal2001,
kang2002}. The high-order statistics of clustering, e.g. the three-point
correlation and the bispectrum, should be quite sensitive to the
non-sphericity.
(iii) Dynamics of galactic satellites. Recently this has been argued to
be very sensitive to the non-sphericity of the host halo
\citep[e.g.,][]{Ibata01}. 
The combination of those three approaches would even yield a direct test
of the cold dark matter paradigm \citep{spergel00,yoshida00}.  

Of course the non-sphericity of dark matter halos is critical to
understanding that of the gas density profile of clusters of galaxies.
Since gas dynamics is characterized by the isotropic pressure tensor and
does not directly follow the dark matter distribution in halos.  In
fact, most hydrodynamic simulations of galaxy clusters suggest that the
gas distribution is generally rounder than that of dark matter.
Nevertheless we would like to mention a couple of important examples
where the non-sphericity in the gas distribution has crucial and
observable consequences.  (iv) the gas and temperature profiles of X-ray
clusters. Almost all previous analytical models for the X-ray profiles
of galaxy clusters have adopted the spherical approximation perhaps due
to the lack of any specific model for the non-sphericity. Since our
triaxial model specifies the gravitational potential of the hosting
halos, one may compute the gas or temperature profiles, with an
additional assumption of the hydrostatic equilibrium for instance, as
performed in the NFW model \citep[e.g.,][]{MSS98,SSM98,KS01}.  If
combined with the observed surface brightness distribution of clusters,
one may in principle solve for the gas and temperature profiles
simultaneously for a given non-spherical distribution of dark matter
\citep{SW78,YS99,Zaroubi}.  (v) the systematic bias and statistical
distribution of the Hubble constant estimated via the Sunyaev --
Zel'dovich effect. In view of the on-going observational projects, it is
of vital importance to re-evaluate the reliability of the estimates
taking account of the non-sphericity effect of the clusters. With the
above modeling of the gas and temperature profiles for individual
clusters, one may discuss the statistical properties of the estimates of
the Hubble constant\citep[e.g.,][]{FP02}, combining the extensive
fitting formula for the probability distribution functions of the
triaxial model parameters and the halo mass function
\citep[e.g.,][]{sheth99,jenkins01}.

%%%%%%%%%%%%%%%%%%%%%%%%%%%%%%%%%%%%%%%%%%%%%%%%%%%%%%%%%%%%%%%
\acknowledgments
%%%%%%%%%%%%%%%%%%%%%%%%%%%%%%%%%%%%%%%%%%%%%%%%%%%%%%%%%%%%%%%

We thank an anonymous referee for helpful comments on the earlier
manuscript.  Numerical simulations presented in this paper were carried
out at ADAC (the Astronomical Data Analysis Center) of the National
Astronomical Observatory, Japan, and at KEK (High Energy Accelerator
Research Organization, Japan).  Y.P.J. was supported in part by the
One-Hundred-Talent Program, by NKBRSF (G19990754) and by NSFC
(No.10125314), and Y.S was supported in part by the Grant-in-Aid from
Monbu-Kagakusho (07CE2002, 12640231), and by the Supercomputer Project
(No.00-63) of KEK.

\clearpage
%%%%%%%%%%%%%%%%%%%%%%%%%%%%%%%%%%%%%%%%%%%%%%%%%%%%%%%%%%%%%%%%%%%%%%%

%%%%%%%%%%%%%%%%%%%%%%%%%%%%%%%%%%%%%%%%%%%%%%%%%%%%%%%%%%%%%%%%%%%%%%%

\clearpage

%%%%%%%%%%%%%%%%%%%%%%%%%%%%%%%%%%%%%%%%%%%%%%%%%%%%%%%%%%%%%%%%%%%%%%%
\begin{deluxetable}{lccccccccc}
\tablecolumns{9}
\tablewidth{0pc}
\tablecaption{Model parameters for cosmological
simulations with $N=512^3$ in a $100h^{-1}$Mpc box.
 \label{table:cosmosim}}
\tablehead{
\colhead{Model} & \colhead{$\Omega_0$} &  \colhead{$\lambda_0$}
& \colhead{$\sigma_8$}& \colhead{$\Gamma$} 
& $m_p [\himsun]$ & $\eta$ [$h^{-1}$kpc] & timesteps & \colhead{realizations}}
\startdata
LCDM & 0.3  & 0.7 &0.9 & 0.2 & $6.2\times 10^{8}$ & 20 &1200&2\\
SCDM & 1.0  & 0.0 &0.55 & 0.5 & $2.1\times 10^{9}$ & 20 &1200 & 2\\
LCDMa & 0.3  & 0.7 &0.9&0.2 &$6.2\times 10^{8}$ & 10 &5000&1\\
\enddata
\end{deluxetable}
%%%%%%%%%%%%%%%%%%%%%%%%%%%%%%%%%%%%%%%%%%%%%%%%%%%%%%%%%%%%%%%%%%%%%%%

%%%%%%%%%%%%%%%%%%%%%%%%%%%%%%%%%%%%%%%%%%%%%%%%%%%%%%%%%%%%%%%%%%%%%
\begin{deluxetable}{cccccccccccccc}
%\tablefontsize{\footnotesize}
\tablecaption{Properties of the new simulated halos in the LCDM model
 with
$\Omega_0=0.3$, $\lambda_0=0.7$, $h=0.7$, $\sigma_8=1$, and $\Gamma=0.21$}
\tablewidth{0pt}
\tablehead{
\colhead{identification number}
&\colhead{$M [h^{-1} M_\odot]$ \tablenotemark{a}}
&\colhead{$N_p$ \tablenotemark{b}}
&\colhead{$r_{\rm vir} [h^{-1}$Mpc]\tablenotemark{c}}
}
\startdata
GX 5 & $6.1\times 10^{12}$ & 945864 & 0.373\\
GR 5 & $5.5\times 10^{13}$ & 644839 & 0.776 \\
\enddata
\tablenotetext{a}{Mass of the halo within its virial radius.}
\tablenotetext{b}{Number of particles within its virial radius.}
\tablenotetext{c}{the virial radius of the halo.}
\label{table:halosim}
\end{deluxetable}
%%%%%%%%%%%%%%%%%%%%%%%%%%%%%%%%%%%%%%%%%%%%%%%%%%%%%%%%%%%%%%%%%%%%%
\clearpage

%%%%%%%%%%%%%%%%%%%%%%%%%%%%%%%%%%%%%%%%%%%%%%%%%%%%%%%%%%%%%%%
%%%%%%%%%%%%%%%%%%%%%%%%%%%%%%%%%%%%%%%%%%%%%%%%%%%%%%%%%%%%%%%

%%%%%%%%%%%%%%%%%%%%%%%%%%%%%%%%%%%%%%%%%%%%%%%%%%%%%%%%%%%%%%%
%%%%%%%%%%%%%%%%%%%%%%%%%%%%%%%%%%%%%%%%%%%%%%%%%%%%%%%%%%%%%%%

\clearpage

%%%%%%%%%%%%%%%%%%%%%%%%%%%%%%%%%%%%%%%%%%%%%%%%%%%%%%%%%%%%%%%
%%%%%%%%%%%%%%%%%%%%%%%%%%%%%%%%%%%%%%%%%%%%%%%%%%%%%%%%%%%%%%%

%%%%%%%%%%%%%%%%%%%%%%%%%%%%%%%%%%%%%%%%%%%%%%%%%%%%%%%%%%%%%%%
%%%%%%%%%%%%%%%%%%%%%%%%%%%%%%%%%%%%%%%%%%%%%%%%%%%%%%%%%%%%%%%

%%%%%%%%%%%%%%%%%%%%%%%%%%%%%%%%%%%%%%%%%%%%%%%%%%%%%%%%%%%%%%%
%%%%%%%%%%%%%%%%%%%%%%%%%%%%%%%%%%%%%%%%%%%%%%%%%%%%%%%%%%%%%%%

%%%%%%%%%%%%%%%%%%%%%%%%%%%%%%%%%%%%%%%%%%%%%%%%%%%%%%%%%%%%%%%
%%%%%%%%%%%%%%%%%%%%%%%%%%%%%%%%%%%%%%%%%%%%%%%%%%%%%%%%%%%%%%%

%%%%%%%%%%%%%%%%%%%%%%%%%%%%%%%%%%%%%%%%%%%%%%%%%%%%%%%%%%%%%%%
%%%%%%%%%%%%%%%%%%%%%%%%%%%%%%%%%%%%%%%%%%%%%%%%%%%%%%%%%%%%%%%

%\clearpage
%%%%%%%%%%%%%%%%%%%%%%%%%%%%%%%%%%%%%%%%%%%%%%%%%%%%%%%%%%%%%%%
%\begin{figure}[htb]
%\begin{center}
% \leavevmode\epsfxsize=12.0cm \epsfbox{newfig1.eps}
%\end{center} 
%\figcaption{ Correlation of the axis ratios $a/c$ (top), $a/b$ (middle)
%and $c_e$ (bottom) against the virial mass of the halos.  In each panel,
%the results of all halos from the cosmological simulations at $z=0$, 0.5
%and 1.0 are plotted.  The left and right panels correspond to the
%results of LCDM and SCDM models, respectively.  \label{fig:acabcmass} }
%\end{figure}
%%%%%%%%%%%%%%%%%%%%%%%%%%%%%%%%%%%%%%%%%%%%%%%%%%%%%%%%%%%%%%%

%\clearpage
%%%%%%%%%%%%%%%%%%%%%%%%%%%%%%%%%%%%%%%%%%%%%%%%%%%%%%%%%%%%%%%
%\begin{figure}[htb]
%\begin{center}
% \leavevmode\epsfxsize=12.0cm \epsfbox{newfig2.eps}
%\end{center} 
%\figcaption{ Correlation between the axis ratios $a/c$ and $a/b$.  Upper
%and Lower panels correspond to the results of LCDM and SCDM models,
%respectively at $z=0$ (left), 0.5 (center), and 1.0 (right).
%\label{fig:acab} }
%\end{figure}
%%%%%%%%%%%%%%%%%%%%%%%%%%%%%%%%%%%%%%%%%%%%%%%%%%%%%%%%%%%%%%%

\clearpage
%%%%%%%%%%%%%%%%%%%%%%%%%%%%%%%%%%%%%%%%%%%%%%%%%%%%%%%%%%%%%%%
%%%%%%%%%%%%%%%%%%%%%%%%%%%%%%%%%%%%%%%%%%%%%%%%%%%%%%%%%%%%%%%

\clearpage

%%%%%%%%%%%%%%%%%%%%%%%%%%%%%%%%%%%%%%%%%%%%%%%%%%%%%%%%%%%%%%%
%%%%%%%%%%%%%%%%%%%%%%%%%%%%%%%%%%%%%%%%%%%%%%%%%%%%%%%%%%%%%%%

%%%%%%%%%%%%%%%%%%%%%%%%%%%%%%%%%%%%%%%%%%%%%%%%%%%%%%%%%%%%%%%
%%%%%%%%%%%%%%%%%%%%%%%%%%%%%%%%%%%%%%%%%%%%%%%%%%%%%%%%%%%%%%%

%%%%%%%%%%%%%%%%%%%%%%%%%%%%%%%%%%%%%%%%%%%%%%%%%%%%%%%%%%%%%%%
%%%%%%%%%%%%%%%%%%%%%%%%%%%%%%%%%%%%%%%%%%%%%%%%%%%%%%%%%%%%%%%

%%%%%%%%%%%%%%%%%%%%%%%%%%%%%%%%%%%%%%%%%%%%%%%%%%%%%%%%%%%%%%%
%%%%%%%%%%%%%%%%%%%%%%%%%%%%%%%%%%%%%%%%%%%%%%%%%%%%%%%%%%%%%%%

%%%%%%%%%%%%%%%%%%%%%%%%%%%%%%%%%%%%%%%%%%%%%%%%%%%%%%%%%%%%%%%
%%%%%%%%%%%%%%%%%%%%%%%%%%%%%%%%%%%%%%%%%%%%%%%%%%%%%%%%%%%%%%%

%%%%%%%%%%%%%%%%%%%%%%%%%%%%%%%%%%%%%%%%%%%%%%%%%%%%%%%%%%%%%%%
%%%%%%%%%%%%%%%%%%%%%%%%%%%%%%%%%%%%%%%%%%%%%%%%%%%%%%%%%%%%%%%

%%%%%%%%%%%%%%%%%%%%%%%%%%%%%%%%%%%%%%%%%%%%%%%%%%%%%%%%%%%%%%%
%%%%%%%%%%%%%%%%%%%%%%%%%%%%%%%%%%%%%%%%%%%%%%%%%%%%%%%%%%%%%%%

%%%%%%%%%%%%%%%%%%%%%%%%%%%%%%%%%%%%%%%%%%%%%%%%%%%%%%%%%%%%%%%
%%%%%%%%%%%%%%%%%%%%%%%%%%%%%%%%%%%%%%%%%%%%%%%%%%%%%%%%%%%%%%%


\begin{thebibliography}{}
\bibitem[Bardeen et al.(1986)]{BBKS1986} Bardeen, J.M., 
Bond, J.R., Kaiser, N., \& Szalay, A.S. 1986, \apj, 304,15
\bibitem[Bartelmann et al.(1998)]{bart98} 
Bartelmann, M., Huss, A., Colberg, J.~M., Jenkins, A., 
\& Pearce, F.~R.\ 1998, \aap, 330, 1
\bibitem[Barnes \& Efstathiou(1987)]{BE87}
Barnes, J. \& Efstathiou, G. 1987, \apj, 319, 575
\bibitem[Birkinshaw, Hughes \& Arnaud(1991)]{BHA91}
Birkinshaw, M.,  Hughes, J. P., \& Arnaud, K. A. 1991, \apj, 379, 466
\bibitem[Bryan \& Norman(1998)]{BN98}
Bryan, G.L, \& Norman, M., 1998, ApJ, 495, 80
\bibitem[Bullock et al.(2001)]{bullock01}
Bullock, J. S., Kolatt, T. S., Sigad, Y., Somerville, R. S., 
Kravtsov, A. V., Klypin, A. A., Primack, J. R., \& Dekel, A. 2001, 
\mnras, 321, 559
\bibitem[Bullock(2001)]{bullock01b}
Bullock, J. S. preprint astro-ph 0106380
\bibitem[Buote \& Xu(1997)]{bx97} 
Buote, D.~A.~\& Xu, G.\ 1997, \mnras, 284, 439
\bibitem[Davis et al.(1985)]{davis85} 
Davis, M., Efstathiou, G., Frenk, C. S., 
\& White, S. D. M., 1985, ApJ, 292, 371
\bibitem[Dubinski (1994)]{dubinski94}
 Dubinski, J., 1994, ApJ,  431, 617
\bibitem[Efstathiou et al.(1985)]{efstathiou85}
Efstathiou, G., Davis, M., Frenk, C.~S., \& White,
S.~D.~M. 1985, ApJS, 57, 241
\bibitem[Eke, Navarro \& Steinmetz(2001)]{eke01}
Eke, V. R., Navarro, J.F., \& Steinmetz, M. 2001, \apj, 554, 114
\bibitem[Fox \& Pen(2002)]{FP02}
  Fox, D.C., \& Pen, U. 2002, ApJ, submitted (astro-ph/0110311)
\bibitem[Fukushige \& Makino(1997)]{fukushige97}
  Fukushige, T., \& Makino, J. 1997, ApJ, 477, L9
\bibitem[Fukushige \& Makino(2001)]{fukushige01}
  Fukushige, T., \& Makino, J. 2001, \apj, 557, 533
\bibitem[Hamana et al. (2001)]{Hamanaetal2001}Hamana, T., Yoshida, N., 
Suto, Y., \& Evrard, A. E. 2001, ApJ, 561, 143
\bibitem[Hernquist \& Katz(1989)]{HK89}
Hernquist, L. \& Katz, N. 1989, \apjs,70,419
\bibitem[Hockney \& Eastwood (1981)]{HE1981}
Hockney, R. W., \& Eastwood, J. W. 1981,
Computer Simulation Using Particles (McGraw Hill, New York)
\bibitem[Ibata et al.(2001)]{Ibata01}
 Ibata, R. et al.  2001, ApJ, 551, 294
\bibitem[Inagaki, Suginohara \& Suto(1995)]{inagaki95} 
Inagaki, Y., Suginohara, T. \& Suto, Y. 1995, PASJ, 47, 411
\bibitem[Jenkins et al.(2001)]{jenkins01}Jenkins, A., Frenk, C.S.,
 White, S.D.M., Colberg, J.M., Cole, S., Evrard, A.E., Couchman,
 H.M.P. \& Yoshida, N., 2001, MNRAS, 321, 372
\bibitem[Jing(2000)]{jing00}Jing, Y. P. 2000, \apj, 535, 30
\bibitem[Jing \& Fang(1994)]{JF1994} 
 Jing, Y.P. \& Fang, L.Z. 1994,ApJ, 432, 438
\bibitem[Jing, et al. (1995)]{jing95} Jing, Y.~P., Mo, H.~J., 
Borner, G., \& Fang, L.~Z.\ 1995, \mnras, 276, 417
\bibitem[Jing \& Suto(1998)]{JS1998} 
  Jing, Y.P. \& Suto, Y. 1998, ApJ, 494, L5
\bibitem[Jing \& Suto(2000)]{jingsuto00}
Jing, Y. P., \& Suto, Y. 2000, \apjl, 529, L69
\bibitem[Kang et al.(2002)]{kang2002} Kang, X., Jing, Y.P., Mo, H.J., 
\& B\"orner, G., 2002, astro-ph/0201124
\bibitem[Keeton \& Madau(2001)]{keeton01}
Keeton, C. R., \& Madau, P. 2001, \apjl, 549, L25
\bibitem[Kitayama \& Suto(1997)]{kitayama97} 
  Kitayama, T. \& Suto, Y. 1997, ApJ, 490, 557
\bibitem[Klypin, et al. (2001)]{k2001} Klypin, A., Kravtsov, A.~V., Bullock, J.~S., \& Primack, J.~R.\ 2001, \apj, 554, 903
\bibitem[Komatsu \& Seljak(2001)]{KS01} 
  Komatsu, E. \& Seljak, U. 2001, MNRAS, 327, 1353
\bibitem[Lacey \& Cole(1994)]{LC94} 
  Lacey, C. \& Cole, S. 1994, MNRAS, 271, 676
\bibitem[Lahav et al.(2002)]{Lahav02} 
  Lahav, O. et al. 2002, MNRAS, submitted (astro-ph/0112162)
\bibitem[Ma \& Fry (2000)]{MF2000} Ma, C. \& Fry, J.\ N.\ 2000 ApJ, 531, L87
\bibitem[Makino, Sasaki \& Suto(1998)]{MSS98} 
Makino, N., Sasaki, S., \& Suto, Y. 1998, \apj, 497, 555
\bibitem[Meneghetti et al.(2000)]{meneghetti00}
Meneghetti, M., Bolzonella, M., Bartelmann, M., 
Moscardini, L., \& Tormen, G. 2000, \mnras, 314, 338
\bibitem[Meneghetti et al.(2001)]{meneghetti01}
Meneghetti, M., Yoshida, N., Bartelmann, M., Moscardini, L., 
Springel, V., Tormen, G., \& White S. D. M. 2001,  \mnras, 325, 435
\bibitem[Mo, Jing, \& B\"orner (1997)]{MJB1997} Mo, H. J., Jing,Y. P., 
\& B\"orner G. 1997, MNRAS, 286, 979
\bibitem[Molikawa \& Hattori(2001)]{molikawa01}
Molikawa, K., \& Hattori, M. 2001, \apj, 559, 544
\bibitem[Moore et al.(1998)]{moore98}
  Moore, B., Governato, F., Quinn, T., Stadel, J., \& Lake, G. 1998, 
  ApJ, 499, L5
\bibitem[Moore et al.(1999)]{Moore1999}
  Moore, B., Quinn, T., Governato, F., Stadel, J., \& Lake, G. 1999, 
  MNRAS, 310, 1147
\bibitem[Navarro, Frenk, \& White(1996)]{nfw96}
Navarro, J.F., Frenk, C.S., \& White, S. D. M. 1996, \apj, 462, 563
\bibitem[Navarro, Frenk, \& White(1997)]{nfw97}
Navarro, J.F., Frenk, C.S., \& White, S. D. M. 1997, \apj, 490, 493
\bibitem[Oguri(2002)]{oguri02}
Oguri, M. 2002, ApJ, in press (astro-ph/0203142)
\bibitem[Oguri, Taruya, \& Suto(2001)]{oguri01}
Oguri, M., Taruya, A., \& Suto, Y. 2001, \apj, 559, 572
\bibitem[Seljak(2002)]{Seljak02}
Seljak, U. 2002, MNRAS, submitted (astro-ph/0111362)
\bibitem[Sheth \& Tormen(1999)]{sheth99}
Sheth, R. K., \& Tormen, G. 1999, \mnras, 308, 119
\bibitem[Silk \& White(1978)]{SW78}
Silk, J., \& White, S.D.M. 1978, ApJ, 226, L103
\bibitem[Spergel \& Steinhardt(2000)]{spergel00}
 Spergel, D. N., \& Steinhardt P. J. 2000, \prl, 84, 3760
\bibitem[Suginohara \& Suto(1992)]{SS92}
Suginohara, T., \& Suto, Y. 1992, \apj, 396, 395
\bibitem[Suto, Cen \& Ostriker(1992)]{SCO92}
Suto, Y., Cen, R.Y., \& Ostriker, J.P. 1992, \apj, 395, 1
\bibitem[Suto, Sasaki \& Makino(1998)]{SSM98}
Suto, Y., Sasaki, S., \& Makino, N. 1998, \apj, 509, 544
\bibitem[Thomas et al.(1998)]{thomas98} 
 Thomas, P.~A.~et al.\ 1998, \mnras, 296, 1061
\bibitem[Warren et al.(1992)]{Warren92}
Warren, M. S., Quinn, P.J., Salmon, J.K. \& Zurek, W. H. 1992, 
\apj, 399, 405
\bibitem[Yoshida et al.(2000)]{yoshida00}
Yoshida, N., Springel, V., White, S.D.M., \& Tormen, G. 2000, 
\apj, 544, L87
\bibitem[Yoshikawa, Itoh \& Suto(1998)]{yoshikawa98} 
Yoshikawa, K., Itoh, M. \& Suto, Y. 1998, PASJ, 50, 203
\bibitem[Yoshikawa \& Suto(1999)]{YS99} 
Yoshikawa, K., \& Suto, Y. 1999, ApJ, 513, 549
\bibitem[Zaroubi et al.(1998)]{Zaroubi} 
Zaroubi, S., Squires, G., Hoffman, Y., \& Silk, J. 1998, ApJ, 500, L87
\end{thebibliography}
\end{document}